\documentclass[12pt]{article}
\pdfoutput=1

\usepackage{slashed}
\usepackage{color, verbatim}
\usepackage{latexsym}
\usepackage{amsmath}
\usepackage{graphicx}
\graphicspath{{Plots/}}
\usepackage{slashed}
\usepackage{float}

\usepackage{cite}
\usepackage{hyperref}
\usepackage{amssymb}
\usepackage{graphicx}
\usepackage{bm}
\usepackage{hyperref}
\usepackage{adjustbox}
\usepackage{cancel}
\usepackage{soul}

\makeatletter

\@addtoreset{equation}{section}
\makeatother

\usepackage[extra]{tipa}

\begin{document}
\thispagestyle{empty}

\begin{center}

\begin{center}

\vspace{.5cm}

{\Large\sc Composite dark matter phenomenology in the presence of lighter degrees of freedom}

\end{center}

\vspace{0.8cm}

\textbf{
Maria Ramos}\\

\vspace{1.cm}
{\em Laborat\'orio de Instrumenta\c{c}\~ao e F\'isica Experimental de Part\'iculas, Departamento de F\'isica
da Universidade do Minho, Campus de Gualtar, 4710-057 Braga, Portugal}

\end{center}

\begin{abstract}
Scalar singlet dark matter in anomaly-free composite Higgs models is 
accompanied by exotic
particles to which the dark matter annihilates. 
The latter can therefore 
freeze
out even in the absence of couplings to the Standard Model. In this regime, both current and future direct detection constraints can be avoided. Moreover, due to the different decay modes 
of the extra particles, the dark matter candidate can even escape indirect detection
constraints. Assessing this issue requires dedicated simulations of the gamma ray spectrum, that we provide in the present article in the context of $SO(7)/SO(6)$.
For the parameter space region that evades constraints from dark matter experiments, we develop new analyses to be performed at a future 100 TeV collider based on the search of the new particles produced in the decay of heavy vector-like quarks.

\end{abstract}

\newpage

\tableofcontents

 \section{Introduction}
% %
Scalar dark matter (DM) in the context of composite Higgs models (CHM)~\cite{Dimopoulos:1981xc,Kaplan:1983fs,Kaplan:1983sm} has received important attention in the last years~\cite{Frigerio:2012uc, Chala:2012af, Marzocca:2014msa, Fonseca:2015gva, Carmona:2015haa, Ma:2015gra, Bruggisser:2016ixa, Ballesteros:2017xeg, Balkin:2017aep, Ma:2017vzm, Balkin:2017yns, Balkin:2018tma, Chala:2018qdf, Alanne:2018xli, Jiang:2019soj, Davoli:2019tpx, Cacciapaglia:2019ixa, DaRold:2019ccj, DiLuzio:2019wsw, Ruhdorfer:2019utl}. The reason is that both the Higgs boson and the DM are composite pseudo Nambu-Goldstone bosons (pNGBs) of the \textit{same} new strongly interacting sector with global symmetry breaking pattern $\mathcal{G}/\mathcal{H}$ at a scale $f \sim {\rm TeV}$. Therefore, not only their masses are protected against large radiative corrections, but also their size is expected to be of the same order of magnitude, following the WIMP paradigm: the evidence that a weakly interacting massive particle, whose abundance is set by thermal freeze-out \cite{PhysRevLett.39.165}, can account for the whole relic abundance of DM. 

From the model building perspective, CHMs can be also more predictive than elementary DM frameworks. In particular, the symmetries of the strong sector and the way they are broken to generate the scalar potential constrain the number of free parameters at low-energy. It has been shown \cite{Chala:2018qdf} that, by forcing the explicit symmetry breaking to preserve a $\mathcal{Z}_2$ symmetry, one can actually assure \textit{(i)} the stability of DM even after electroweak symmetry breaking (EWSB) and \textit{(ii)} the smallness of the portal coupling to the Higgs boson, as required from current DM constraints. Furthermore, the DM candidate can annihilate sufficiently, avoiding the overclose of the Universe. This is due to new derivative interactions, that emerge after integrating the strong sector out, and which do not appear in the elementary models. 

The smallest coset giving a DM singlet scalar $\eta$ on top of the Higgs doublet $H$ is $SO(6)/SO(5)$~\cite{Gripaios:2009pe}. The corresponding DM phenomenology was first studied in Ref.~\cite{Frigerio:2012uc}. Likewise, $SO(5)\times U(1)/SO(4)$~\cite{Gripaios:2016mmi}, despite being non-simple, has a similar spectrum. In these minimal models, the symmetry stabilizing DM is assumed to be respected both classically \textit{and} at the quantum level; if only the first assumption is made, terms such as $\eta B_{\mu\nu} B^{\mu\nu}$ with $B$ the $U(1)_Y$ field-strength tensor can arise, spoiling the DM stability. In non-anomalous CHMs with DM, it is sufficient to assume that the stabilizing symmetry is respected at the classical level; then it is automatically preserved at the quantum level as well. All CHMs with DM satisfying this feature contain further scalar degrees of freedom that modify the DM phenomenology.
The smallest model relies on $SO(7)/SO(6)$~\cite{Chala:2016ykx,Balkin:2017aep,Balkin:2018tma,DaRold:2019ccj}. Other models with additional singlets include \textit{e.g.} $SO(8)/SO(7)$, $SO(5)/SO(4)\times SU(2)/U(1)$,  $SU(6)/SO(6)$. These extra pNGBs have been assumed heavier than the DM candidate in previous works.
There are however no strong arguments in favor of this assumption.  
Moreover, if a CHM with DM is to explain also other particle physics puzzles like electroweak (EW) baryogenesis via a strong first-order EW phase transition \cite{Chala:2016ykx}, it must actually contain new (scalar) degrees of freedom at or below the EW scale.

Relying on different well-motivated versions of $O(7)/O(6)$, we will demonstrate that the mass $m$ of the extra pNGB can be naturally lighter than the DM mass $m_{DM}$. 
Therefore, the DM does not annihilate only into the SM particles (Higgs and gauge bosons mostly, since other decays are suppressed by explicit symmetry breaking terms). Instead, it annihilates also into the new scalar. 

In this case, the corresponding amplitude scales with $\sim [\lambda - 4m_{DM}^2/f^2]$, where $\lambda$ is the portal coupling to the DM. The SM $\lambda$ arises in the leading-order term of the radiatively induced potential. Therefore, naive power counting arguments~\cite{Giudice:2007fh, Chala:2017sjk} tell us that ${m_{DM}\sim g_* f/(4\pi)}$ with $g_*$ the typical strong coupling between composite resonances; while $\lambda\sim g_*^2/(4\pi)^2 \sim m_{DM}^2/f^2$, partially canceling each other in the annihilation cross section. On the other hand, the $\lambda$ of the new singlet arises often only at next-to-leading order in the expansion of the Coleman-Weinberg potential on the symmetry breaking parameters, its size being therefore $\sim 1/g_*^2$ smaller. Consequently, the new scalar can dominate the DM annihilation. (In models with several new scalars, such as ${SO(N+1)/SO(N)}$ with $N\gg 1$, this result holds trivially irrespective of the aforementioned cancellation.) We substantiate these arguments in sections~\ref{sec:generic} and ~\ref{sec:relic}.

As a corollary, the DM can freeze out even in the absence of couplings to the SM. We show in section~\ref{sec:direct} that, in this case, the current very strong direct detection constraints can be avoided. We also discuss the reach of future experiments to DM-nucleon cross sections induced by loops involving the new scalars. Likewise, depending on how the new scalar decays, the DM can even escape current and future indirect detection constraints. We study these limits in section~\ref{sec:indirect}, performing dedicated simulations of the resulting gamma ray spectrum, since the DM annihilation final state contains non-SM particles for which the experimental collaborations do not provide bounds. We present for the first time the upper bounds from Fermi-LAT on the annihilation cross section, for a fixed $m$, that allows the pNGB DM to account for the whole relic abundance. The results can be straightforwardly recast to other non-minimal cosets. 

Finally, we develop new analyses aimed to unravel the pNGBs at colliders in section~\ref{sec:colliders}. This discussion is significantly important in our work, since a wide region of the parameter space is unconstrained by the DM searches. We study the non-minimality of the model focusing on three decay channels of the extra singlet: $\gamma \gamma$, $b \overline{b}$ and $\mu^+\mu^-$. We provide prospects for a future $pp$ collider, running at a center of mass energy (c.m.e.) $\sqrt{s} = 100$ TeV.
We conclude in section~\ref{sec:conclusions}. 

\section{Generic Lagrangian}\label{sec:generic}

The coset $SO(7)/SO(6)$ delivers two pNGBs in addition to the Higgs fields, full singlets of the SM gauge group, that we will denote by $\eta$ (the DM) and $\kappa$, respectively. The coset generators can be chosen to be:
\begin{align}
 T_{ij}^{mn} &= -\frac{\text{i}}{\sqrt{2}}(\delta^m_i\delta^n_j - \delta^n_i\delta^m_j)~, \quad m<n\in[1,6]~,\\
 X_{ij}^{m7} &= -\frac{\text{i}}{\sqrt{2}}(\delta^m_i\delta^7_j - \delta^7_i\delta^m_j)~, \quad m\in[1,6]~.
\end{align}
The operators $X^{17}-X^{67}$ expand the coset space, while $T^{mn}$ are unbroken.
Without loss of generality \cite{Gripaios:2009pe}, the Goldstone matrix reads
\begin{equation}
 U = 	\left[\begin{array}{cccccc}
 \mathbf{1}_{3\times 3} & & & \\[0.1cm]
 					 & 1 - h^2/(f^2 + \Sigma) & 
-h \eta/(f^2 + \Sigma) & -h \kappa/(f^2 + \Sigma) & 
h/f \\[0.1cm]
 					 & -h\eta/(f^2 + \Sigma) & 1-  
\eta^2/(f^2 + \Sigma) &- \eta \kappa/(f^2 + \Sigma) & 
\eta/f \\[0.1cm]
 					 &- h\kappa/(f^2 + \Sigma) & - 
\eta \kappa/(f^2 + \Sigma) & 1 -  \kappa^2/(f^2 + \Sigma) & 
\kappa/f \\[0.1cm]
 					 & - h/f & - \eta/f & 
-\kappa/f & \Sigma/f^2 \\
	\end{array}\right]~,
\end{equation}
with $\Sigma = f^2 (1-h^2/f^2 -\eta^2/f^2 -\kappa^2/f^2)^{1/2}$.
In the unitary gauge and before EWSB, the Lagrangian consists of a shift-symmetry preserving part described by the nonlinear sigma model
\begin{align}
  L & = -\frac{1}{4}f^2 \text{Tr} \bigg[(U^T\partial_\mu U)_X (U^T\partial^\mu U)_X\bigg] + \mathcal{O}(\partial^4)\\
&  =\frac{1}{2}\bigg[ (\partial_\mu h)^2 + (\partial_\mu \eta)^2 + (\partial_\mu \kappa)^2\bigg]  + \frac{1}{2 f^2} \bigg[(\eta \partial_\mu \eta)^2 + (\kappa \partial_\mu \kappa)^2 + (h \partial_\mu h)^2\bigg] \nonumber\\
 & + \frac{1}{f^2}\bigg[(\eta\partial_\mu\eta)( h\partial^\mu h + 
\kappa\partial^\mu\kappa )  + (\kappa\partial_\mu \kappa) (h \partial^\mu h)  \bigg] +\mathcal{O}(1/f^4) \nonumber~;
\end{align}
where the sub-index $X$ stands for projection over the broken generators; and a shift-symmetry breaking part involving both the Yukawa Lagrangian and the potential:
\begin{equation}
L_{int} =-\frac{y_q}{\sqrt{2}} \overline{q_L} q_R h\bigg[1 +i c_\kappa  \frac{\kappa}{f} - c_\eta \frac{\eta^2}{2 f^2} + ...  \bigg]  + h.c. - V (\eta, \kappa, h)~,\\
\label{eq:Ly}
\end{equation}
where
\begin{equation}
V(\eta, \kappa, h) = \frac{1}{2}m_\eta^2 \eta^2 + \frac{1}{2}m_\kappa^2\kappa^2 + \frac{1}{4}\lambda_{\eta H} \eta^2 h^2 + \frac{1}{4}\lambda_{\eta\kappa} \eta^2 \kappa^2~
 +  \frac{1}{4}\lambda_{\kappa H} \kappa^2 h^2+\cdots~
\label{eq:V}
\end{equation}
The ellipsis stands for other terms not relevant for the DM
phenomenology, including the Higgs potential, $V_H =\mu_H^2h^2/2 + \lambda_H h^4/4$. Equation \ref{eq:Ly} holds similarly for leptons. 
The interactions with the fermions and the potential arise at tree level and one loop, respectively, after integrating the strong sector out. Being shift-symmetry breaking, they depend on the quantum numbers of the composite resonances the elementary fermions mix with, according to the partial compositeness paradigm \cite{KAPLAN1991259}.  We will restrict to mixings that preserve the DM stability $\eta \rightarrow -\eta$, namely we will require that the parity transformation $P_\eta = \rm{diag} (1,1,1,1,-1,1,1)$ be a symmetry of the composite sector and not spontaneously broken. Being this coset anomaly-free, $P_\eta$ can indeed remain unbroken at the quantum level. We 
will further require that the model parameters are not in tension with experimental bounds on CP violation. The most general embeddings for the SM fermions compatible with these assumptions give opposite CP-parities for $\eta$ and $\kappa$: we assume that $\eta$ is a scalar and $\kappa$ is a pseudoscalar, with $c_{\eta, \kappa} \in \mathbb{R}$ in equation \ref{eq:Ly}. We will also assume that these couplings are family universal, in order to avoid bounds from flavour changing neutral currents \cite{Gripaios:2009pe}.

The smallest representations of $SO(7)$ under which the composite resonances can transform are $\mathbf{1}$, $\mathbf{7}$, $\mathbf{21}$ and $\mathbf{27}$. We will consider two cases, in which $q_L\oplus t_R$ mix with operators in the $\mathbf{27}\oplus \mathbf{1}$ and $\mathbf{7} \oplus \mathbf{7}$, respectively. As we will discuss in the subsequent sections, these cases can lead to the following values of the free parameters in equations \ref{eq:Ly} and \ref{eq:V}:
\begin{align}
 \text{RegI}: ~ &\lambda_{\eta H} \sim \lambda_H, ~ \lambda_{\eta\kappa} \ll 
1, ~ f\sim \frac{m_\eta}{\sqrt{\lambda_H}}\left[1 + \frac{m_\kappa^2}{2 m_\eta^2}\right]^{1/2}~;\label{eq:regI}\\[0.2cm]
 \text{RegII}: ~ &\lambda_{\eta H} \ll 1, ~ \lambda_{\eta\kappa}\sim\lambda_H, 
~ f\sim 1, 2.5, 3, 4~\text{TeV}~,
\label{eq:regII}
\end{align}
for $m_\kappa \le m_\eta$. 
In RegII, the benchmark values for $f$ are those compatible with EW precision constraints \cite{Ghosh:2015wiz} and \textit{natural} solutions to the hierarchy problem. (The Higgs mass fine-tuning scales approximately as $\sim m_H^2 /f^2$ \cite{Panico:2015jxa} which is already $\lesssim 10^{-3}$ for $f > 4$ TeV.)

\subsection{Regime I}
The most general embedding of $q_L$ in the representation $\mathbf{27}$ compatible with $P_\eta$
can be described by the spurion
\begin{equation}
\label{eq:QL_t}
 \text{Q}_L = b_L\Lambda^b + t_L \Lambda^t = 	
\frac{1}{\sqrt{2}}\left(\begin{array}{ccc}
              0_{5\times 5} & \gamma\mathbf{v_1}^T & \mathbf{v_2}^T\\
              \gamma\mathbf{v_1} & 0 & 0\\
              \mathbf{v_2} & 0 & 0
             \end{array}\right)~,
\end{equation}
with $\mathbf{v_1} = (b_L, -ib_L, t_L, it_L, 0)$,  $\mathbf{v_2} = (ib_L, b_L, 
it_L, -t_L, 0)$ and $\gamma \geq 0$. For $\gamma = 0$, the singlet $\kappa$ becomes also stable. In this case, $(\eta,\kappa)$ transforms as a complex scalar under a $SO(2) \cong U(1)$ that belongs to the unbroken $SO(6)$ symmetry group.
The corresponding phenomenology was studied in Ref. \cite{Balkin:2018tma}. For $\gamma = 1$, $\kappa$ becomes massless; its shift symmetry remains unbroken. These values are therefore stable under radiative corrections and hence \textit{technically} natural.

The leading-order scalar potential depends only on two unknowns, parameterizing the two non-redundant invariants that can be built out of $\Lambda^{b,t}$:
\begin{align}
\label{V:regI}
 V &= c_1 \bigg[\left(\Lambda_D^{\mathbf{1 *}} \right)^\alpha 
\left(\Lambda_D^{\mathbf{1}} \right)_\alpha \bigg] + 
c_2\bigg[\left(\Lambda_D^{\mathbf{6 *}} \right)^\alpha_i 
\left(\Lambda_D^{\mathbf{6}} \right)^i_\alpha	\bigg]\\\nonumber
 &= 2 c_1\bigg[f^2 h^2 - h^4 - h^2\eta^2 - (1 - \gamma^2) h^2\kappa^2\bigg] \\\nonumber
 & + 4 c_2\bigg[ \frac{1}{4}f^2 (\gamma^2 - 7) h^2 + h^4 - f^2\eta^2 -  
(1-\gamma^2)f^2\kappa^2 + h^2 \eta^2 + (1-\gamma^2)h^2\kappa^2\bigg]~,
\end{align}
where $(\Lambda_D^\alpha)^{\mathbf{1},\mathbf{6}}$ are the projections of 
$U^\dagger \Lambda^\alpha U$ into the singlet and the sextuplet in 
the decomposition $\mathbf{27}=\mathbf{1} \oplus \mathbf{6} \oplus \mathbf{20}$, respectively. Here, $\alpha = b, t$ whereas the index $i$ runs over the generators transforming in the $\mathbf{6}$ representation of $SO(6)$.

Trading $c_1$ and $c_2$ by the Higgs quadratic term $\mu_H$ and its quartic coupling $\lambda_H$, we obtain
\begin{align}
\label{eq: V27}
 V = \frac{\mu_H^2}{2} h^2 + \frac{\lambda_H}{4} h^4 &
 +  \bigg[\frac{\lambda_H f^2 + 2 \mu_H^2}{3 - \gamma^2}\bigg]\eta^2 + \frac{1}{4}\lambda_H \eta^2 h^2 \\
 &+ \bigg[\frac{(\lambda_H f^2 + 2 \mu_H^2)(\gamma^2 - 1)}{\gamma^2 - 
3}\bigg]\kappa^2 + \frac{1}{4}\lambda_H (1-\gamma^2)h^2\kappa^2\nonumber~,
\end{align}
Therefore, $m_{\kappa} \leq m_\eta$ for all values of $\gamma \in [0, 1]$.

On another front, the Yukawa Lagrangian to dimension 6 reads:
\begin{align}
\label{eq:Yuk27}
L_Y &  = y f \overline{q_L^\alpha} \left(\Lambda^\alpha_D \right)^\dagger_{77} t_R + h.c.\\
&  =  - y_q\frac{h}{\sqrt{2}}\overline{t_L} t_R \bigg[ \sqrt{1 - \frac{h^2}{f^2} - \frac{\eta^2}{f^2} - \frac{\kappa^2}{f^2}} + i\gamma \frac{\kappa}{f}\bigg] +h.c. \nonumber\\
&  =  -y_q\frac{h}{\sqrt{2}}\overline{t_L} t_R \bigg[1 + i\gamma \frac{\kappa}{f} - \frac{\eta^2 + \kappa^2 + h^2}{2 f^2}\bigg] + \mathcal{O}\left(\frac{1}{f^4}\right) +  h.c.\nonumber~,
\end{align}
which matches equation \ref{eq:Ly} with $c_\eta = 1$ and $c_\kappa = \gamma$.

For $\gamma \in\mathbb{R}$ and $\left<\eta\right> = \left<\kappa\right> = 0$, both $h$ and $\eta$ can be defined as CP-even scalars, while $\kappa$ is CP-odd, as can be inferred from their couplings to fermions. For $1< \gamma < \sqrt{3}$, we can also have $\left<\kappa\right> \neq 0$, leading to spontaneous CP violation. Since this is very constrained experimentally \cite{Andreev:2018ayy}, in the following we only consider $\gamma$-values that satisfy $\left< \kappa \right> = 0$.  
On the other hand, for $\gamma > \sqrt{3}$, $\eta$ can gain a vacuum expectation value (VEV) and the DM becomes unstable.  Requiring that both CP and $P_\eta$ are not broken spontaneously therefore assures that the aforementioned hierarchy, $m_\kappa \leq m_\eta$, is always respected. This supports our case of study, showing that the traditional assumption that DM is the lightest pNGB in the non-minimal spectrum is not necessarily true.

If we further express the potential in terms of $m_\kappa$ and $m_\eta$, we obtain
finally:
\begin{equation}
 V = \frac{\mu_H^2}{2} h^2 + \frac{\lambda_H}{4} h^4 +  \frac{1}{2} m_\eta^2 \eta^2 + \frac{1}{4}\lambda_H \eta^2 h^2 + \frac{1}{2} m_\kappa^2 \kappa^2 + \frac{1}{4}\lambda_H \left(\frac{m_\kappa}{m_\eta}\right)^2 h^2\kappa^2~,
 \label{eq:Vmasses27}
\end{equation}
that matches equation \ref{eq:regI}.

\subsection{Regime II}
The SM $q_L$ and $t_R$ can be instead both embedded in the fundamental representation $\mathbf{7}$ of $SO(7)$. The embeddings read as follows:
\begin{equation}
\text{T}_R = \left(0,0,0,0,0,i\gamma t_R, t_R\right) \quad \text{and} \quad \text{Q}_L = \frac{1}{\sqrt{2}}\left( -ib_L, b_L, it_L, t_L, 0,0,0\right)~.
\end{equation}
Under $SO(6)$, $\mathbf{7} = \mathbf{1} \oplus \mathbf{6}$; therefore, only one independent invariant can be built at leading order for each field. The leading-order potential reads:
\begin{align}
\label{V:regII}
 V &= c_1 \bigg[\left(\Lambda_{D,R}^{\mathbf{1 *}} \right) 
\left(\Lambda_{D,R}^{\mathbf{1}} \right) \bigg] + 
c_2\bigg[\left(\Lambda_{D,L}^{\mathbf{1 *}} \right)^\alpha
\left(\Lambda_{D,L}^{\mathbf{1}} \right)^\alpha	\bigg]\\\nonumber
 &= c_1\bigg[h^2 f^2 + \eta^2 f^2 + \kappa^2 \left(1 - \gamma^2\right)f^2 \bigg]  + \frac{c_2}{2} h^2 f^2 ~.
\end{align}
One sees clearly that next-to-leading order terms are mandatory to account for EWSB. There are
eight non-redundant invariants that we can build (including the ones already defined in equation \ref{V:regII}):
\begin{align*}
& I_1 \equiv \bigg[\left(\Lambda_{D,R}^{\mathbf{1 *}} \right) 
\left(\Lambda_{D,R}^{\mathbf{1}} \right) \bigg] = f^2 \bigg[f^2 - h^2 - \eta^2 - \kappa^2 \left(1 - \gamma^2\right)  \bigg] \\
&I_2 \equiv \bigg[\left(\Lambda_{D,L}^{\mathbf{1 *}} \right)^\alpha
\left(\Lambda_{D,L}^{\mathbf{1}} \right)^\alpha	\bigg] =  \frac{1}{2}h^2 f^2~;\\
&  I_3 \equiv \bigg[ \left(\Lambda_{D,R}^{\mathbf{1 *}} \right) 
\left(\Lambda_{D,R}^{\mathbf{1}} \right)\bigg]^2 = \bigg[ h^2 + \kappa^2 \left( 1- \gamma^2\right) + \eta^2 - f^2 \bigg]^2~;\\
& I_4 \equiv \bigg[ \left(\Lambda_{D,R}^{\mathbf{6 *}} \right)^i 
\left(\Lambda_{D,R}^{\mathbf{6}} \right)_i \bigg]^2  = \bigg[ h^2 + \kappa^2 \left(1- \gamma^2 \right) + \eta^2 + \gamma^2 f^2 \bigg]^2~;\\
&  I_5 \equiv \bigg[ \left(\Lambda_{D,L}^{\mathbf{1 *}} \right)^\alpha 
\left(\Lambda_{D,L}^{\mathbf{1}} \right)_\alpha \bigg]^2 = \frac{1}{4}h^4~;\\
& I_6 \equiv \bigg[ \left(\Lambda_{D,L}^{\mathbf{6 *}} \right)^\alpha_i 
\left(\Lambda_{D,L}^{\mathbf{6}} \right)_\alpha^i  \bigg]^2 =\frac{1}{4} \left(h^2 - 4 f^2\right)^2~;\\
 & I_7 \equiv  \left(\Lambda_{D,R}^{\mathbf{1 *}} \right) 
\left(\Lambda_{D,R}^{\mathbf{1}} \right) \left(\Lambda_{D,L}^{\mathbf{1 *}} \right)^\alpha 
\left(\Lambda_{D,L}^{\mathbf{1}} \right)_\alpha  = \frac{1}{2}h^2 \bigg[ f^2 - h^2 - \eta^2 - \kappa^2 \left(1-\gamma^2\right)\bigg]~;\\
& I_8 \equiv \left(\Lambda_{D,R}^{\mathbf{6 *}} \right)^i 
\left(\Lambda_{D,R}^{\mathbf{6}} \right)_i \left(\Lambda_{D,L}^{\mathbf{1 *}} \right)^\alpha 
\left(\Lambda_{D,L}^{\mathbf{1}} \right)_\alpha  = \frac{1}{2} h^2 \bigg[ h^2 + \eta^2 + \kappa^2 \left(1-\gamma^2\right) + \gamma^2 f^2 \bigg]~.
\end{align*}
It is clear from the form of  $I_1-I_8$ that the naive counting of nine operators constructed out of $\{h^2, \eta^2, \kappa^2\}$ is reduced to only five, by combining $\{h^2, \eta^2 + (1-\gamma^2) \kappa^2\}$. The corresponding five coefficients plus $\gamma$ can be traded by the Higgs mass and quartic coupling, the two scalar masses and the two quartic couplings.
After this replacement, we finally obtain:
\begin{align}
\label{eq:Vmasses7}
V = & \frac{\mu_H^2}{2} h^2 + \frac{\lambda_H}{4} h^4 + \frac{1}{2} m_\eta^2 \eta^2 + \frac{1}{8} \lambda_{\eta \kappa} \left(\frac{m_\eta}{m_\kappa}\right)^2 \eta^4 + \frac{1}{2} m_\kappa^2 \kappa^2 + \frac{1}{8} \lambda_{\eta \kappa} \left(\frac{m_\kappa}{m_\eta}\right)^2 \kappa^4 \\
& + \frac{1}{4} \lambda_{\eta H} \eta^2 h^2 +  \frac{1}{4} \lambda_{\eta \kappa}\eta^2 \kappa^2 + \frac{1}{4} \lambda_{\eta H} \left(\frac{m_\kappa}{m_\eta}\right)^2 \kappa^2 h^2~.\nonumber
\end{align}

Based on this construction, we conclude that the DM couplings to the singlet and to the Higgs boson can be chosen independently. These two portal couplings are unknown and generated at the same order, in contrast to what was obtained in the previous subsection. In order to study a complementary phenomenology to that of RegI, we assume $\lambda_{\eta H} < \lambda_{\eta \kappa} \sim \lambda_H$. Moreover, we explore the limiting case of $\lambda_{\eta H} \rightarrow 0$ 
\footnote{The suppression $\lambda_{\eta H} \lesssim 0.1 \lambda_{\eta \kappa}$, with $\lambda_{\eta \kappa} \sim \lambda_H$ can be achieved as a result of 1\% level of tuning in the model parameters. (We do a numerical scan varying all the relevant UV constants, uniformly distributed, in $[0.1c,10c]$ where $c\sim 0.1 \lambda_H$ in order to reproduce the Higgs quartic coupling). This (minimal) tuning is comparable to the Higgs mass tuning required to have $v^2 \ll f^2$ in the radiatively induced potential; similar values were obtained in Ref. \cite{Balkin:2017aep}.}
, in order to minimize constraints from direct detection experiments and explore whether indirect detection constraints can be sensitive to this regime.

On the other hand, the Yukawa Lagrangian, $L_Y = y_f f \overline{q_L^\alpha}\left(\Lambda_{D,L}^{\mathbf{1 *}} \right)^\alpha \left(\Lambda_{D,R}^{\mathbf{1}} \right)t_R + h.c.$, gives the same result as equation \ref{eq:Yuk27}.
Despite being less predictive than the previous scenario, this regime also yields the condition $m_\kappa \leq m_\eta$ in the parameter space compatible with our assumptions; see equation \ref{V:regII}.

\begin{figure}[t]
 \includegraphics[width=0.49\columnwidth]{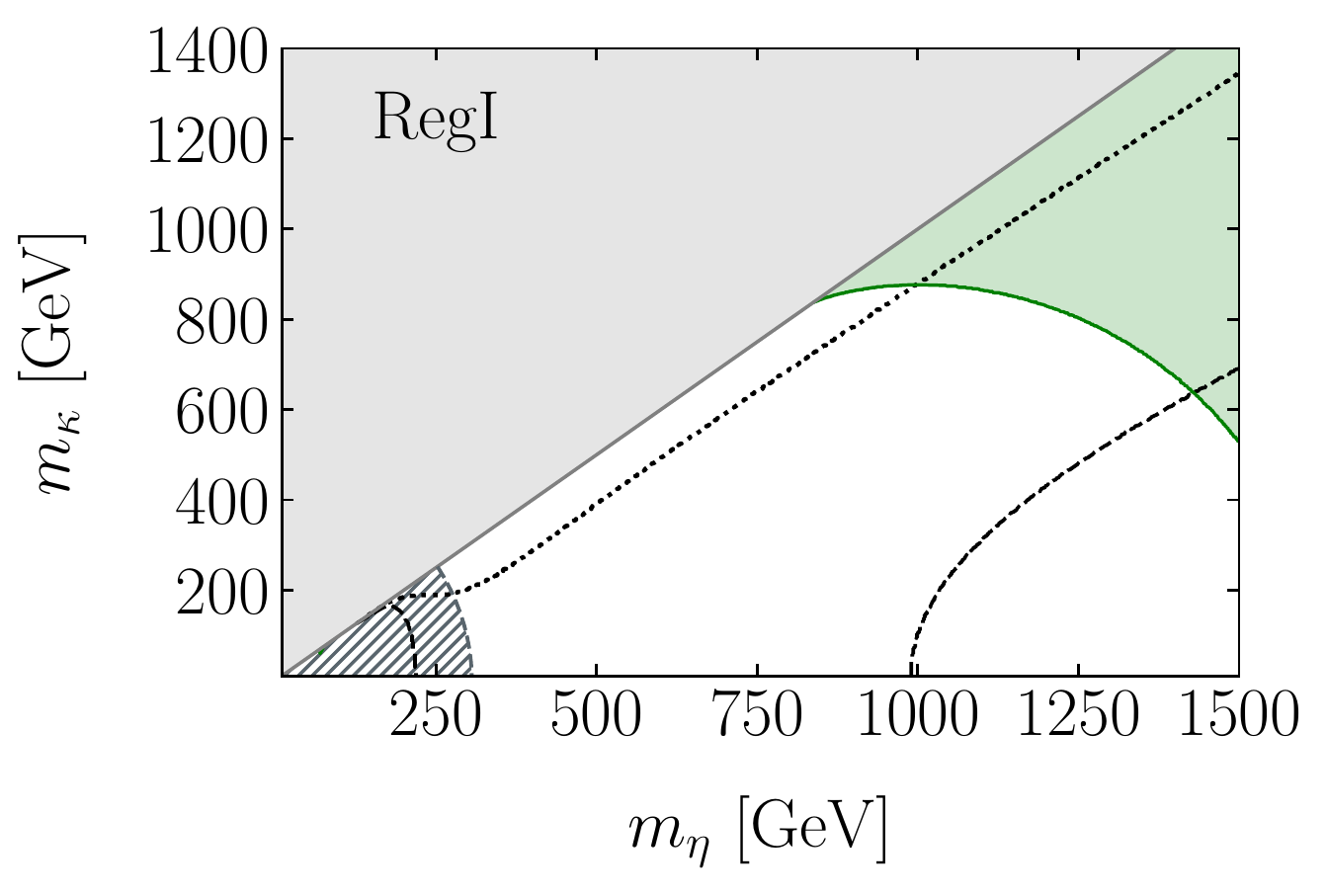}
 \includegraphics[width=0.49\columnwidth]{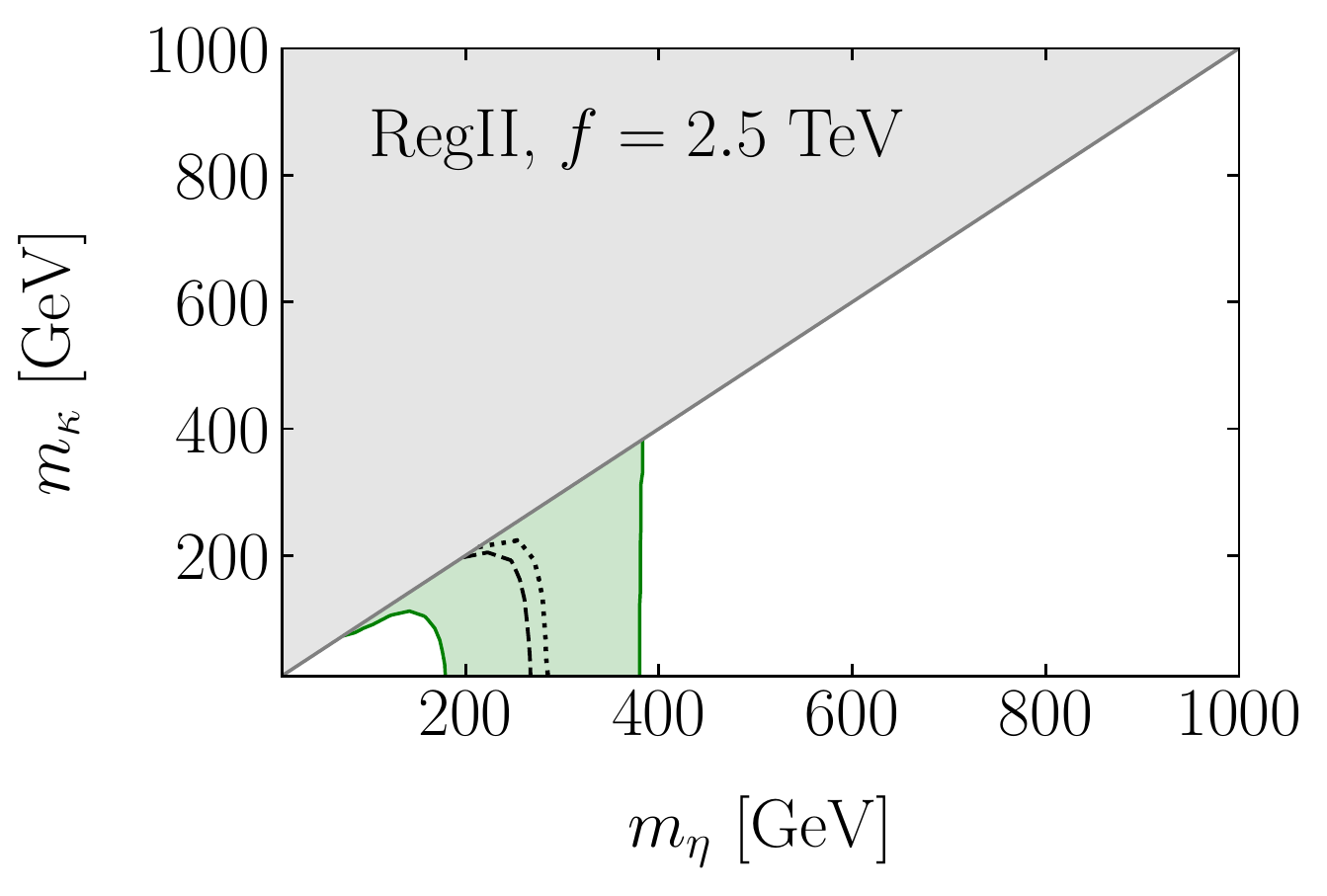}
 \includegraphics[width=0.49\columnwidth]{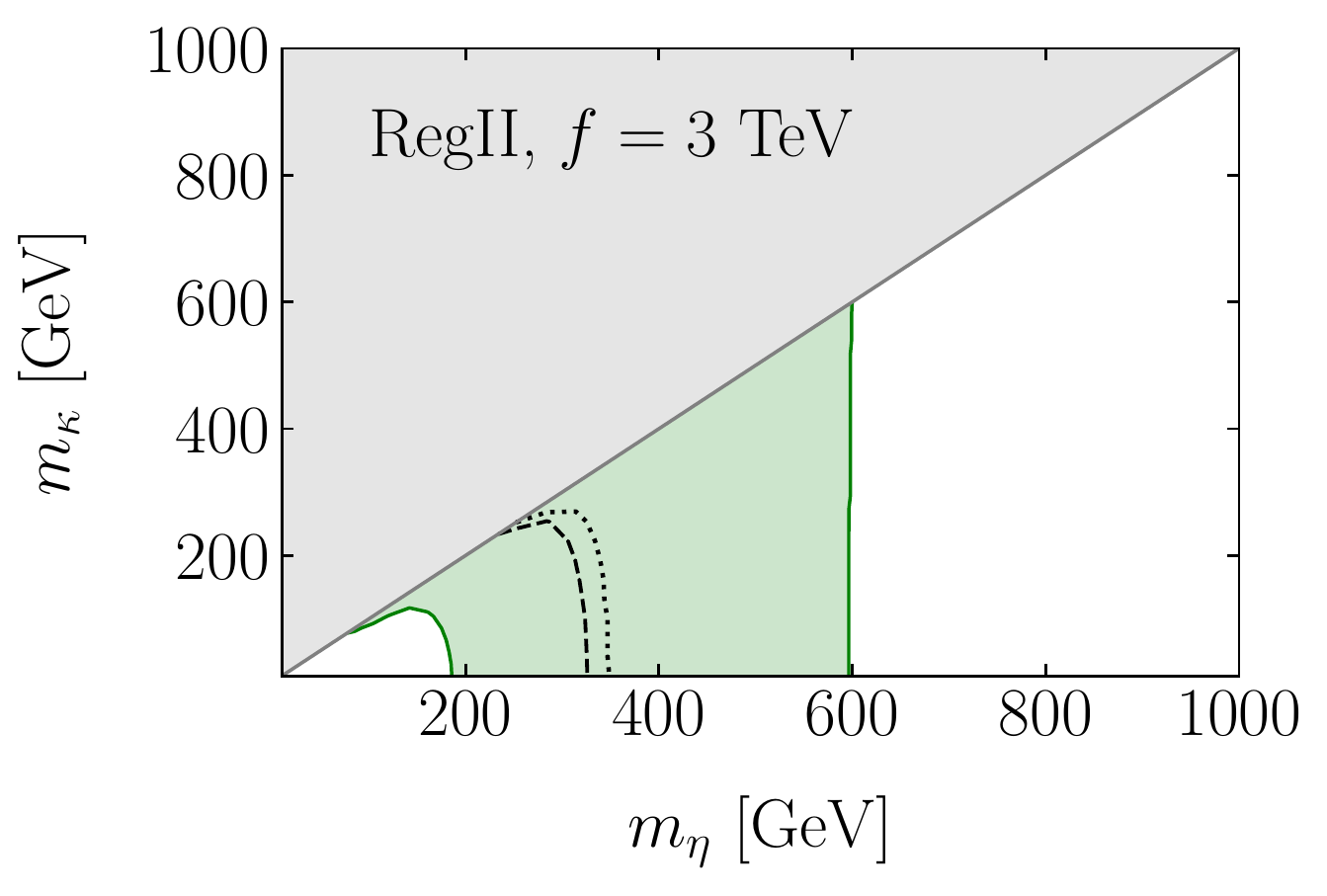}
 \includegraphics[width=0.49\columnwidth]{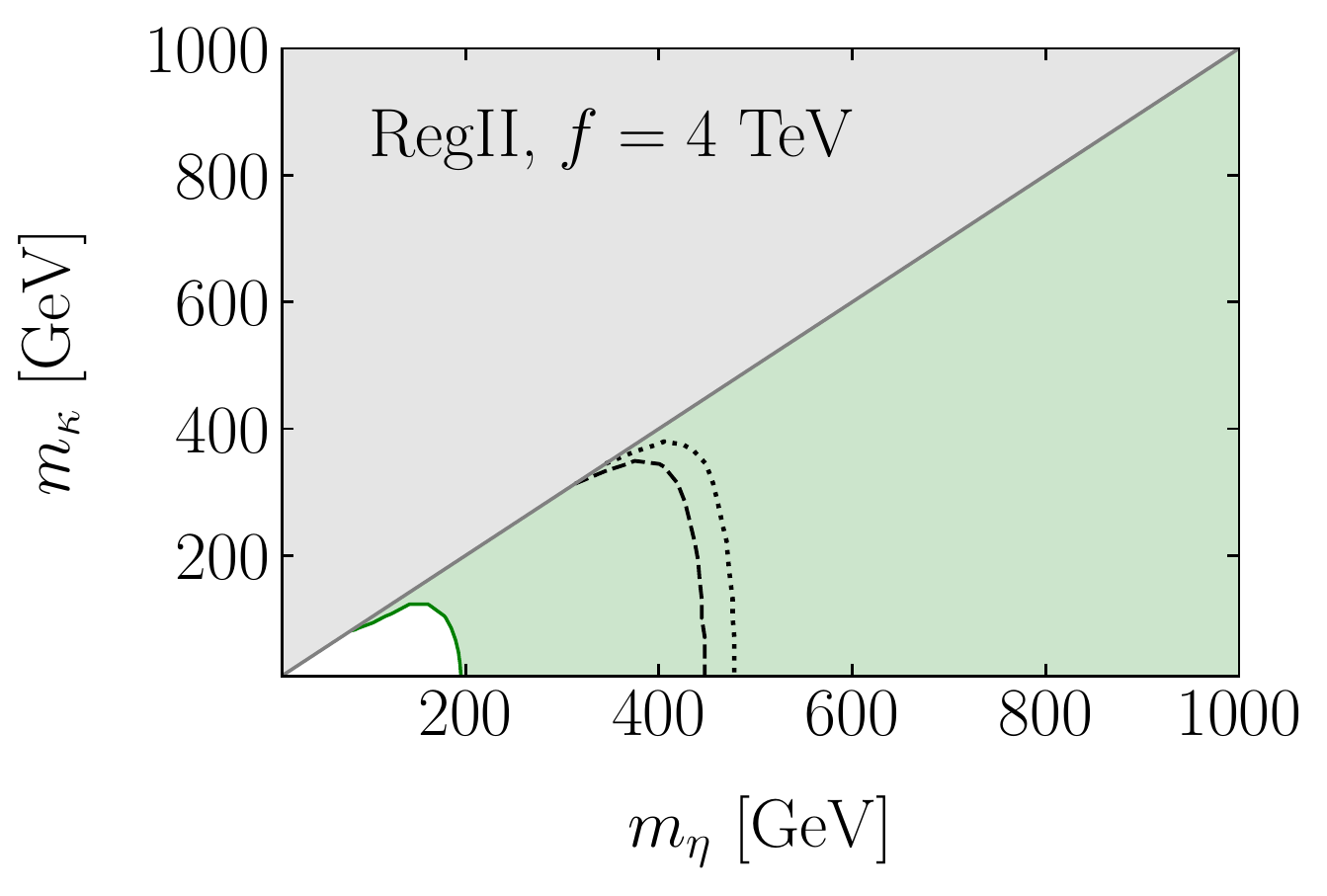}
 \caption{\it In the region enclosed by the solid green line, DM is overabundant if it is a thermal relic. The dotted and dashed black lines correspond to an annihilation fraction into $\kappa \kappa$ equal to $0.2$ and $0.3$, respectively. The area enclosed by the solid gray line is theoretically forbidden. The upper left panel and the upper right, bottom left and bottom right panels stand for Reg I and Reg II with $f = 2.5, 3$ and $4$ TeV, respectively. The DM portal couplings $(\lambda_{\eta H}, \lambda_{\eta \kappa})$ are set to $(\lambda_H, 0)$ and $(0, \lambda_{H})$, in RegI and RegII, respectively. In both regimes, we use $c_\eta = 1$. In RegI, the slashed area is excluded by EWPD \cite{Ghosh:2015wiz}.
 }
 \label{fig:relic}
\end{figure}

\section{Relic density}\label{sec:relic}
The DM annihilation cross section into Higgses in the limit of small DM 
velocity $\omega$ reads
\begin{equation}\label{eq:xsec}
 \sigma \omega (\eta\eta\to h h) \simeq \frac{1}{64\pi m_\eta^2} \bigg[ 
\lambda_{\eta H} - \frac{4 m_\eta^2}{f^2} \bigg]^2 
\bigg[1-\frac{m_h^2}{m_\eta^2}\bigg]^{1/2}~,
\end{equation}
and similarly for the EW gauge bosons by virtue of the Goldstone equivalence 
theorem for $m_\eta \gg v$, with $v \sim 246$ GeV denoting the Higgs VEV; and likewise for $\eta\eta\to\kappa\kappa$ upon the replacements $m_h,\lambda_{\eta H}\to m_\kappa,\lambda_{\eta\kappa}$. 
One easily notes that, in RegI, there is a partial cancellation between the portal coupling to the Higgs boson and the term originating from the derivative interactions, in the first bracket of equation \ref{eq:xsec}. The extra scalar can therefore dominate the DM annihilation. In Reg II, this cancellation occurs instead in $\sigma \omega \left(\eta \eta \rightarrow \kappa \kappa\right)$. After EWSB, the Higgs coupling to the DM is parameterized by $\left(\lambda_{\eta H} - 4m_\eta^2/f^2\right) v$; consequently, the cancellation behavior also occurs in the $s$-channel annihilation into the SM (or $\kappa \kappa$).

To clarify further this point, we determine the region of the plane $(m_\eta, m_\kappa)$ for which the relic density $\Omega h^2$ is above the measured value $\Omega h^2_\text{obs}\sim 0.12$~\cite{Aghanim:2018eyx}. To this aim, we use \texttt{micrOmegas}~\cite{Belanger:2014hqa} with a \texttt{CalcHEP}~\cite{Pukhov:1999gg} model obtained by means of \texttt{Feynrules}~\cite{Christensen:2008py}. The corresponding area is enclosed by the solid green line in figure~\ref{fig:relic} for Reg I and for Reg II with $f = 2.5, 3, 4$ TeV (the parameter space for $f= 1$ TeV is unconstrained). Contour lines of constant annihilation fraction into $\kappa \kappa$ are also plotted.  In RegI, the slashed gray area corresponds to values of $f$ excluded by EW precision data (EWPD).

Let us first focus on the case of RegII and analyze the results for $f=2.5$ TeV. It is clear that the behavior of the relic density can be described in three regions of increasing $m_\eta$: \textit{(i)} For $m_\eta^2 /f^2 \ll 1$, the annihilation cross section into $\kappa$ grows quadratically with $\lambda_{\eta \kappa}\sim \lambda_H\sim 0.1$; it is still large enough for $\Omega h^2 < \Omega h^2_{\rm obs}$. \textit{(ii)} Eventually, $m_\eta^2 / f^2\sim \lambda_{\eta\kappa}$, leading to destructive interference between the two terms in equation \ref{eq:xsec}; therefore, DM becomes overabundant. \textit{(iii)} When $m_\eta^2/f^2$ is large enough, the derivative term dominates the dynamics and the DM abundance decreases once again. In the transition from \textit{(ii)} to \textit{(iii)}, the observed relic abundance is attained quite independently from $m_\kappa$. For a given $f$, the value for $m_\eta$ at which this transition takes place matches the one from Eq. (12) in Ref. \cite{Balkin:2018tma}.

This picture is modified in RegI where $f\propto m_\eta/\sqrt{\lambda_H}$ (see equation \ref{eq:regI}) and, therefore, the derivative interactions are already significant for small DM masses. Concretely:
\begin{equation}
\sigma (\eta\eta\to h h )\propto \frac{\lambda_H^2}{m_\eta^2} \left(1 - \frac{4}{1 + \frac{m_\kappa^2}{2m_\eta^2}}\right)^2~,\quad \sigma(\eta\eta \to \kappa\kappa) \propto \frac{\lambda_{H}^2}{m_\eta^2} \left(\frac{4}{1+\frac{m_\kappa^2}{2m_\eta^2}}\right)^2~.
\end{equation}
Therefore, increasing $m_\kappa$ while keeping $m_\eta$ fixed ($i.e.$ increasing $f$), the effective couplings become smaller and $\eta$ becomes overabundant. Increasing $m_\eta$ instead, while keeping $m_\kappa$ fixed, makes the effective couplings larger; however, as the cross sections scale with $m_\eta^{-2}$, this will eventually lead to overabundance as well.

Let us note that the requirement $\Omega h^2 = \Omega h^2_\text{obs}$ establishes a relation between $m_\eta$ and $m_\kappa$, effectively removing one free parameter. (This conclusion fails obviously if the DM is non-thermal; we will also consider this case in next sections.) In RegI, the corresponding relation reads approximately 
\begin{align}
m_\kappa  \left(m_\eta\right) \approx & \big[ -5461.99 + 21.002 \left(m_\eta/{\rm GeV}\right) -2.6313\times 10^{-2} \left(m_\eta/{\rm GeV}\right)^2 \big. \nonumber \\
& \big. +1.4988\times 10^{-5} \left(m_\eta/{\rm GeV}\right)^3 - 3.3382 \times 10^{-9} \left(m_\eta/{\rm GeV}\right)^4 \big] ~{\rm GeV},
\label{eq:fit}
\end{align}
for $m_\eta \in  [\sim 800, 1500]$ GeV.

\color{black}
Furthermore, in RegI, the relic density constraint sets a bound on $2.8 \lesssim f \lesssim 3.3$ TeV. This contrasts with CHMs without DM, for which this upper bound relies only on fine-tuning arguments, rather than on actual observables.

\section{Direct detection}\label{sec:direct}

\begin{figure}[t]
\begin{center}
 \includegraphics[width=0.8\columnwidth]{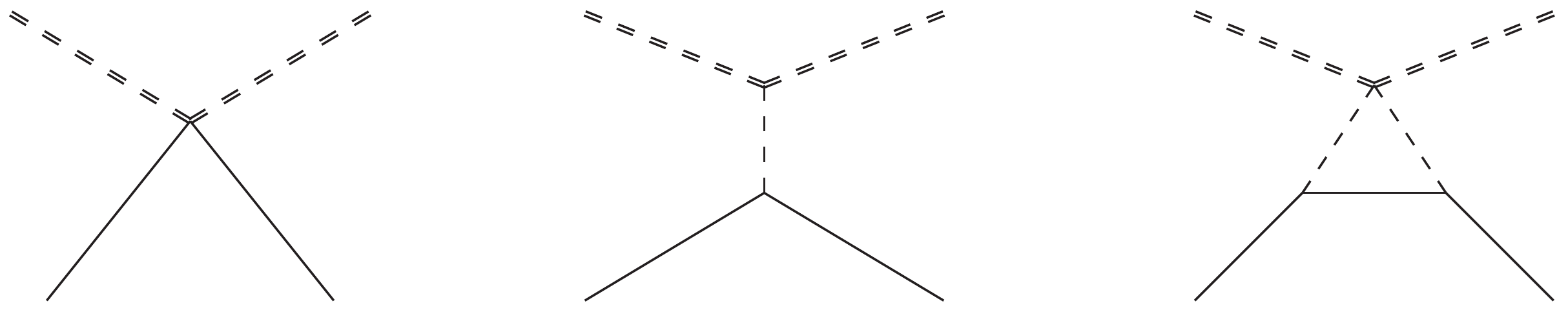}
 \end{center}
 \caption{\it Contributions to the operator $\eta^2\overline{q} q$ at low energies from the contact interaction (left), from the exchange of a Higgs boson in $t$-channel (center) and from a loop of $\kappa$ (right).}\label{fig:diagrams}
\end{figure}
\begin{figure}[ht!]
 \begin{center}
  \includegraphics[width=0.49\columnwidth]{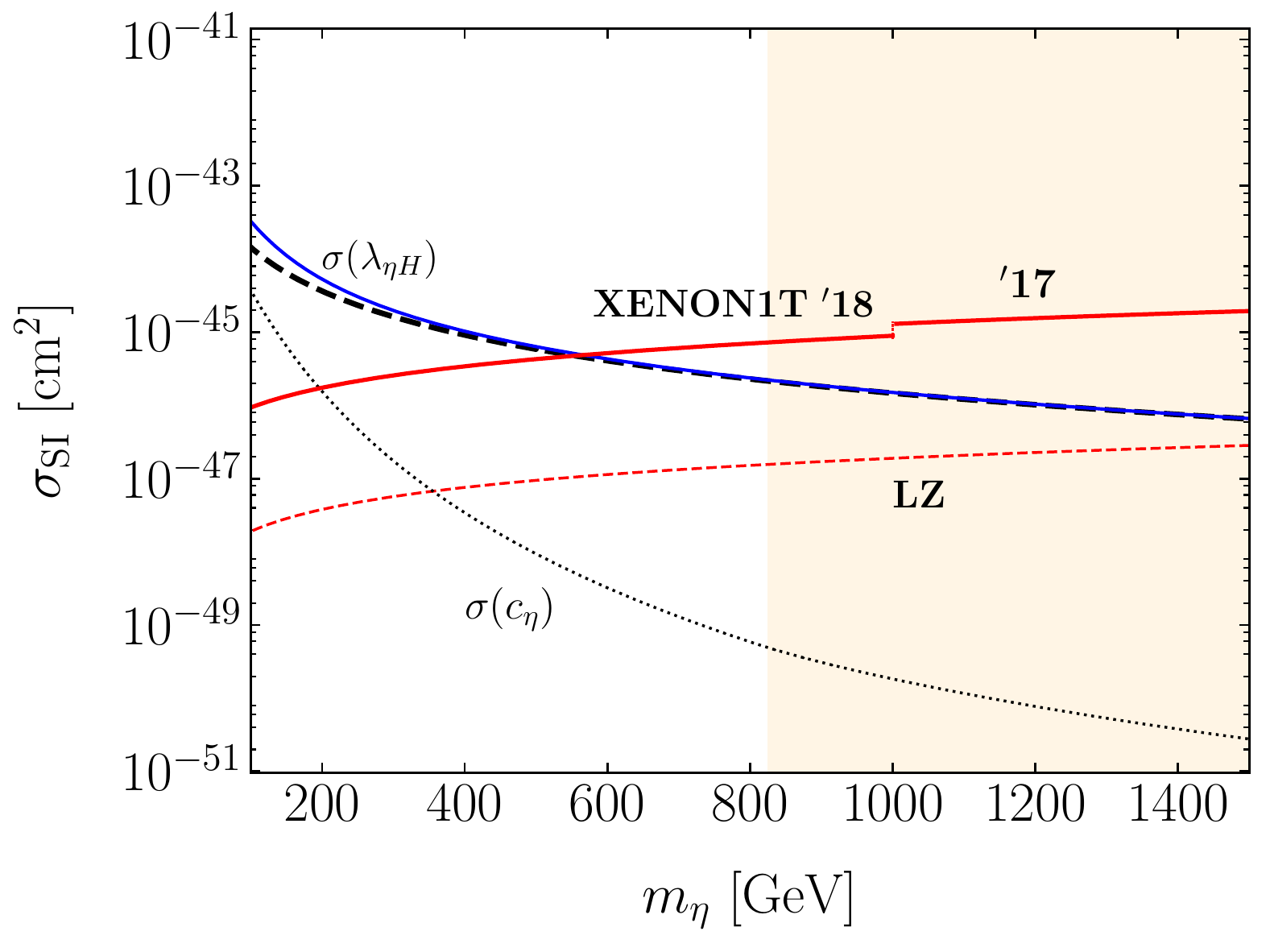}
  \includegraphics[width=0.49\columnwidth]{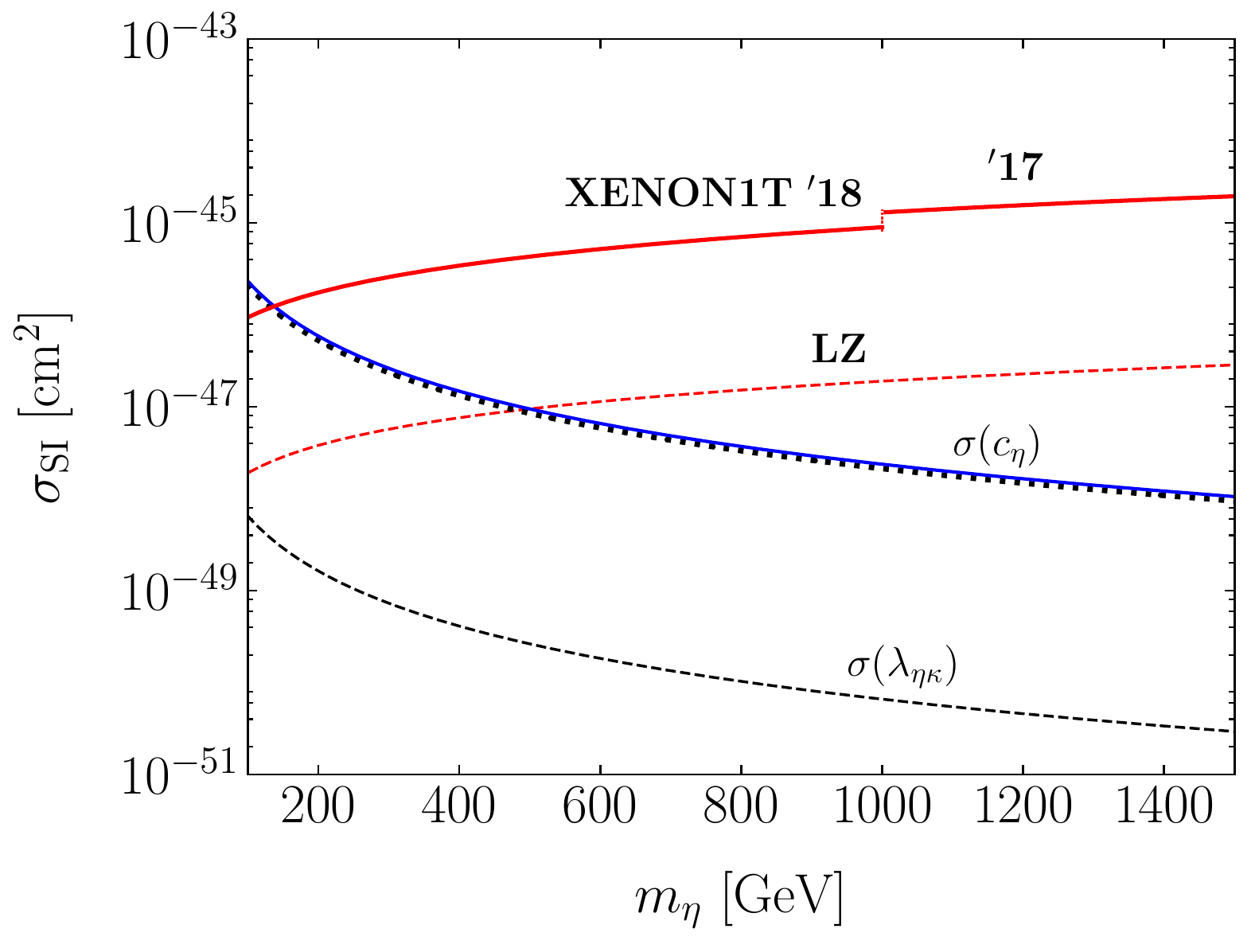}
  \includegraphics[width=0.49\columnwidth]{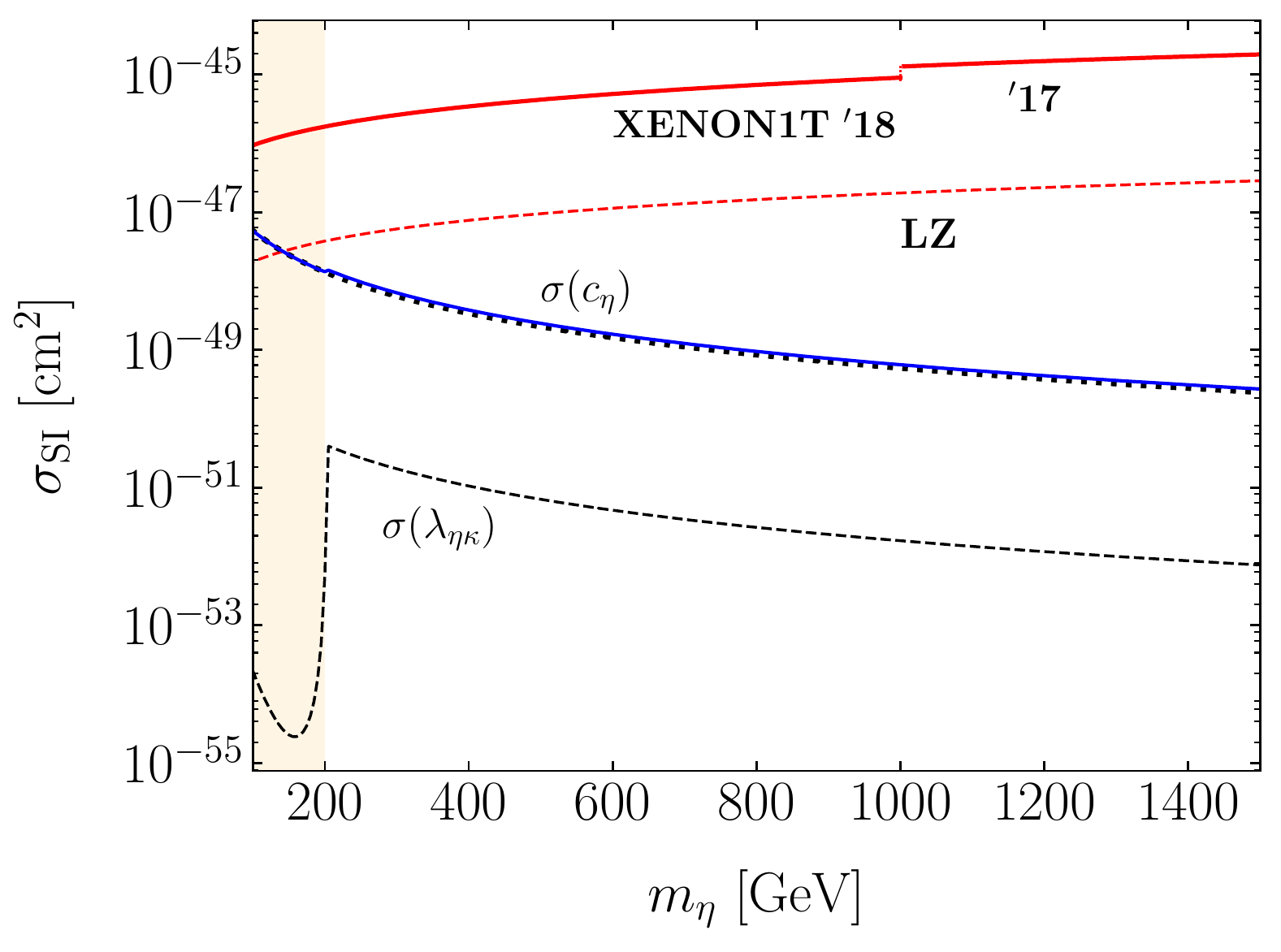}
  \includegraphics[width=0.49\columnwidth]{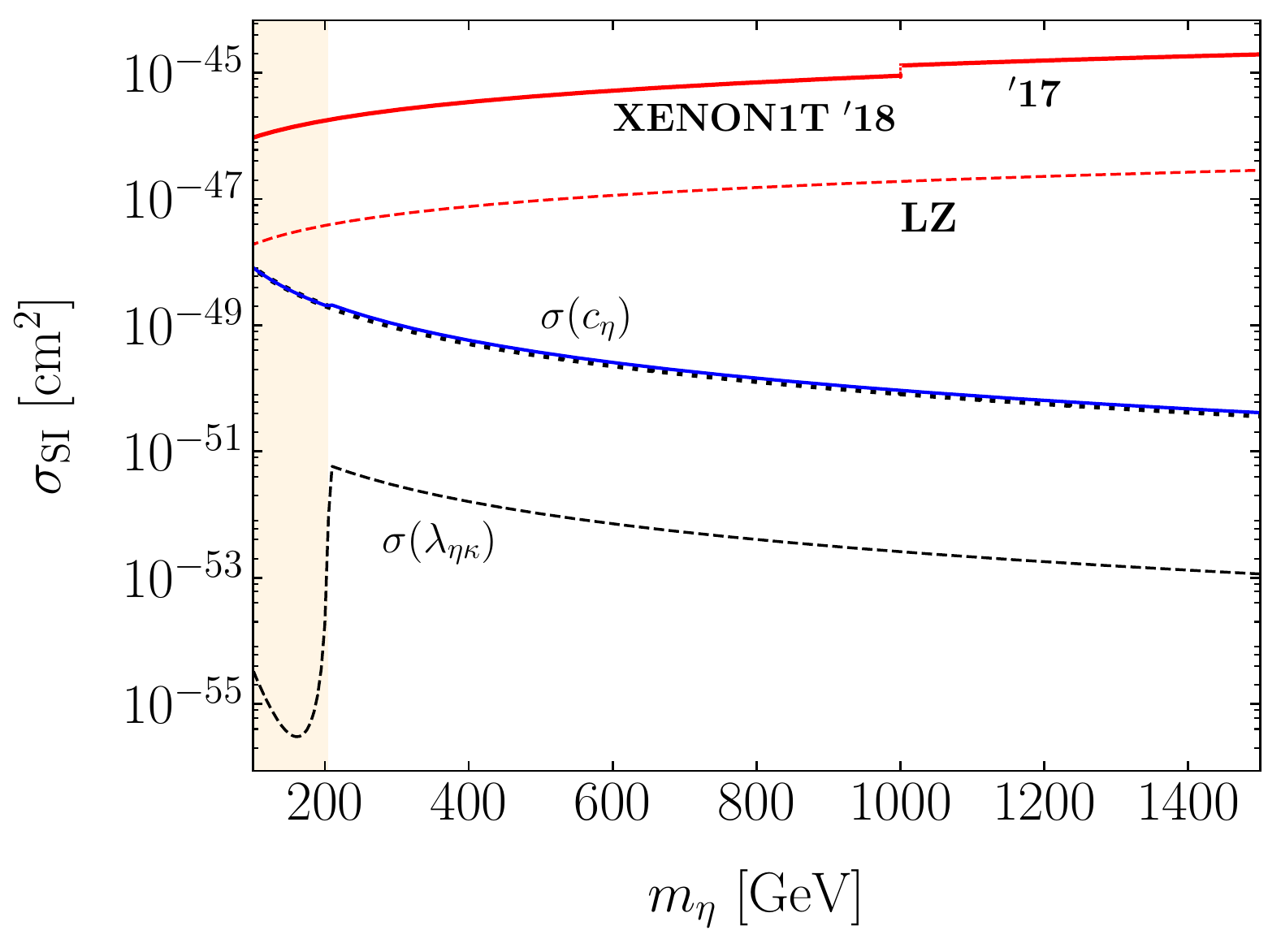}
 \end{center}
 \caption{\it Spin independent cross section for the elastic scattering of DM off nuclei as a function of its mass. In the orange area, $m_\kappa$ is a function of the DM mass to fit $\Omega_{\rm obs} h^2$; see the text for more details. Outside the orange region, it is assumed that all DM is made of only $\eta$. The solid red and dashed curves are for the XENON1T and LZ exclusion limits, respectively. The black dashed and dotted lines stand for the scalar mediated and contact interaction contributions in each regime; whereas the blue line represents the sum of all contributions.
In the upper left panel, the parameters of the model are those of RegI.  In the upper right, bottom left and bottom right panels, we represent the results for RegII with $f = 1,~2.5$ and $4$ TeV.}
 \label{dd_xs}
\end{figure}

Direct detection experiments such as XENON1T~\cite{Aprile:2017iyp} or the future LZ~\cite{Akerib:2018lyp} search for signs of DM in the recoils of atomic nucleus when scattered off by DM particles. These are Earth-based ultra-sensitive experiments with low backgrounds and have set some of the strongest bounds on WIMPs. Both XENON1T and LZ use liquid xenon as a target for direct detection, which is very efficient in converting the low energy from nuclear recoils into observable signals (scintillation and ionization) and highly sensitive to the spin-independent (SI) DM interactions due to its large mass number.

Explicitly, the SI elastic scattering cross section of $\eta$ on a nucleus reads:
\begin{equation}
\sigma_{\rm SI} = \frac{1}{\pi} \bigg(\frac{m_\eta m_n}{m_\eta + m_n}\bigg)^2 
\frac{\left[Z f_p + \left(A-Z\right) f_n \right]^2}{A^2}~,
\end{equation}
where $m_n\sim 1$ GeV is the neutron mass; $Z$ and $A-Z$ are the numbers of protons and 
neutrons in the nucleus, respectively; and the form factors are defined as  \cite{Frigerio:2012uc}:
\begin{equation}
f_{n,p} = \sum_{q = u,d,s} f_{T_q}^{(n,p)} a_q \frac{m_{n,p}}{m_q} + 
\frac{2}{27} f_{T_G}^{(n,p)} \sum_{q=c,b,t} a_q \frac{m_{n,p}}{m_q}~.
\label{eq:ff}
\end{equation}
The matrix elements $f_{T_q}$ parametrize the quark content of the 
nucleon. We consider those from Ref. \cite{Frigerio:2012uc}. The second term in equation \ref{eq:ff} refers to the coupling of DM to the gluons in the nucleus, through a loop of heavy quarks, and is given by $f_{T_G} = 1- \sum_{q=u,d,s} f_{T_q}$. Finally, $a_q$ stands for the coefficient of the effective operator $m_q \eta^2 \overline{q} 
q$ at low energies. In the scenario under consideration, the latter receives tree level contributions from
$c_\eta$ and $\lambda_{\eta H}$; see the left and center diagrams in figure~\ref{fig:diagrams}. If they vanish, loops of $\kappa$ provide the dominant contribution to $a_q$. All together, we obtain \footnote{In order to compute the loop contribution to the scattering cross section, we make use of the approximation that $m_\kappa$ is the largest scale involved in the low-energy nuclear interaction; in particular, $m_\kappa^2 \gg t^2$, where $t$ stands for the Mandelstam variable.}:
\begin{equation}
a_q = \frac{m_q}{m_\eta} \bigg[ \frac{c_\eta}{2 f^2 } + \frac{\lambda_{\eta H}}{2 m_h^2} + \frac{\lambda_{\eta \kappa} c_\kappa^2 }{32\pi^2} \left(\frac{m_q}{ m_\kappa f}\right)^2 \bigg]~.
\label{eq:aq}
\end{equation}
In figure \ref{dd_xs}, we represent each contribution from equation \ref{eq:aq} to the SI scattering cross section, as a function of the DM mass. For each black line, only one coupling is given a non-zero value; we work with $c_{\eta, \kappa} = 1$ and $\lambda=\lambda_{H}$, as defined in each regime of equations \ref{eq:regI} and \ref{eq:regII}. The total SI cross section is plotted in blue, while the current and strongest bounds from XENON1T are represented in the red solid line (we use the results from \cite{Aprile:2018dbl} and \cite{Aprile:2017iyp} to constrain masses up to and above 1 TeV, respectively). The projected  bounds from the LZ experiment are also represented by the red dashed line.
In the orange region, we use the relation $m_\kappa (m_\eta)$ obtained in the fit of section \ref{sec:relic}, to study the scenario where $\eta$ is all the observed DM in the Universe. To probe the non-minimal setup, we fix $m_\kappa = 10$ GeV outside this region.

Note that, while RegI is partially excluded by XENON1T, RegII leads to theoretical cross sections which are typically two (four) orders of magnitude below the LZ projected limit for a non-zero (zero) coupling to fermions. Moreover, in the region where $\Omega h^2 = \Omega_{\rm obs} h^2$ (for $f>1$ TeV), the cross section is much more suppressed. This regime therefore evades all direct detection constraints, both current and future.

\section{Indirect detection}\label{sec:indirect}

\begin{figure}[ht!]
 \begin{center}
  \includegraphics[width=0.49\columnwidth]{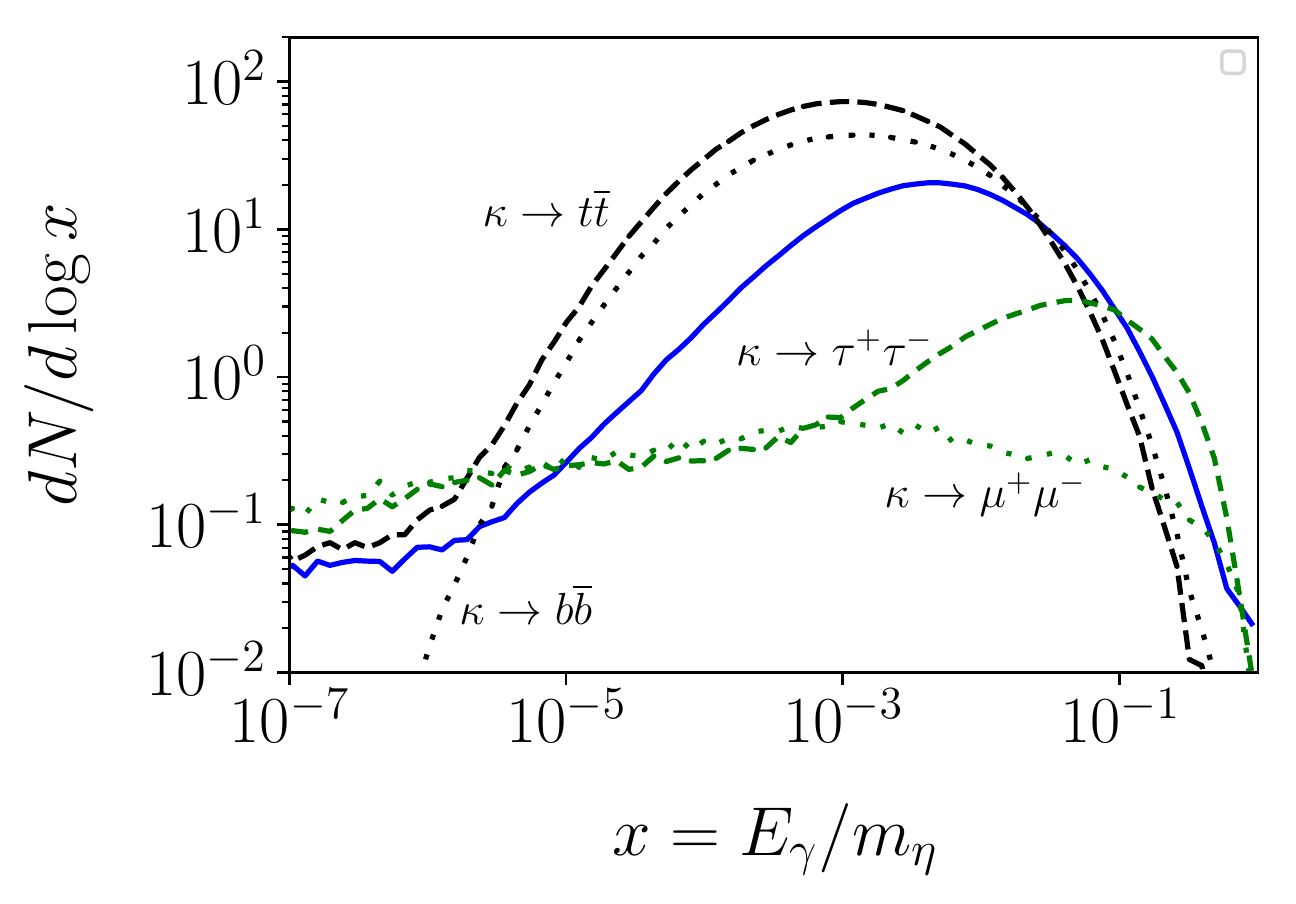}
  \includegraphics[width=0.49\columnwidth]{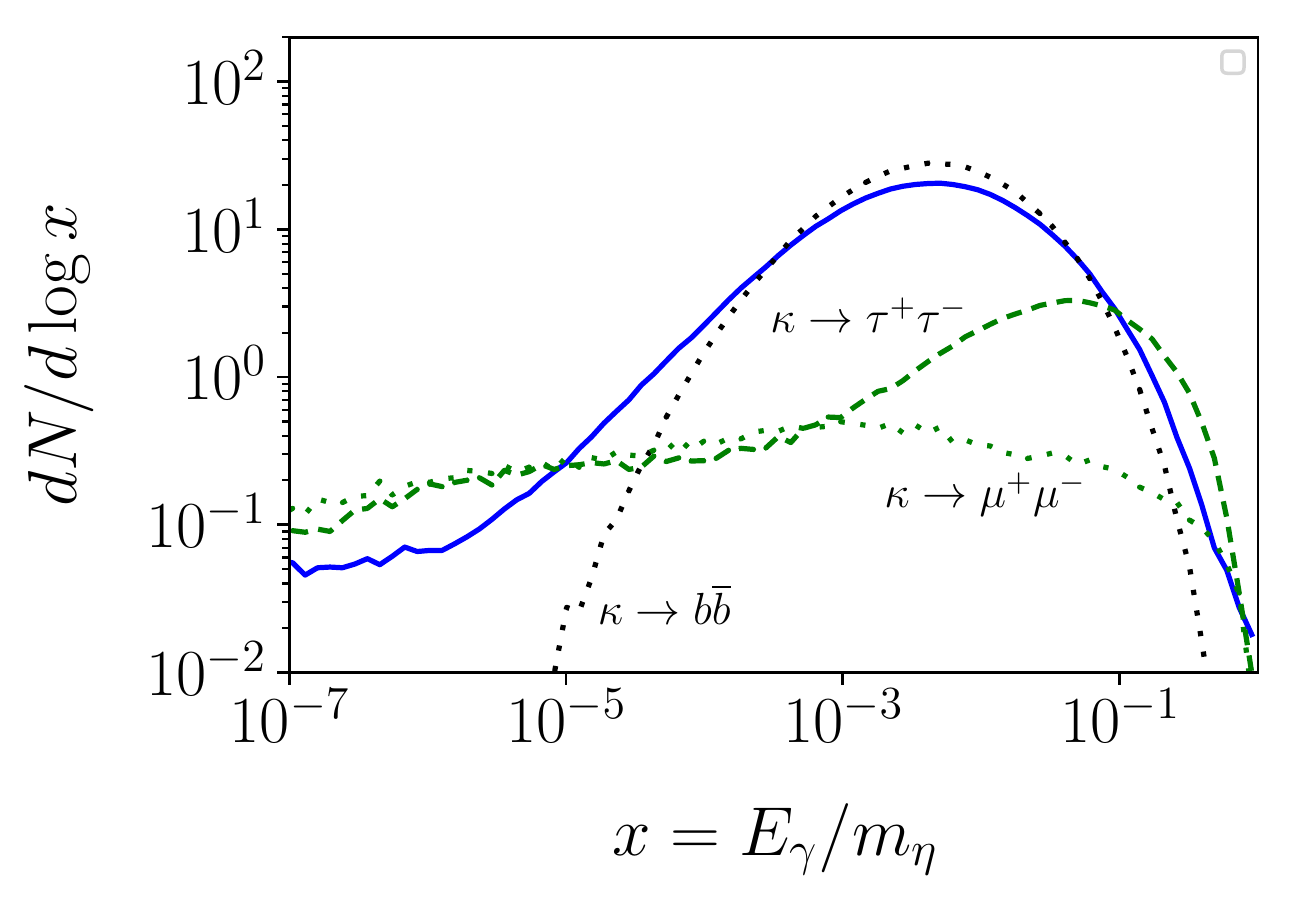}
 \end{center}
\caption{\it Energy spectra $dN/d log{x}$ (with $x = E_\gamma/m_\eta$ and $E_\gamma$ the kinetic energy of the final state photons) of gamma rays at the production point. The blue line corresponds to the DM annihilation into SM bosons, $W^+W^-+ZZ+hh$. The other curves stand for the annihilation into $\kappa\kappa$, each labeling different decay modes of this extra scalar. (The gamma-ray spectrum from $\kappa\rightarrow e^+e^-$ is very similar to that of $\kappa \rightarrow \mu^+ \mu^-$.)  These spectra were computed for a benchmark point defined by $m_\eta = 2 m_\kappa$ in the region where the DM thermal abundance equals the observed value; see equation \ref{eq:fit}. The left panel refers to RegI ($m_\eta \sim 1.3$ TeV); while the right panel is obtained for RegII with $f = 2.5$ TeV ($m_\eta \sim 180$ GeV).}
\label{gammas}
\end{figure}
To obtain the constraints from indirect detection, we resort to \texttt{MadDm v3}~\cite{Ambrogi:2018jqj}, in which we can compute the spectra resulting from DM annihilating into an arbitrary number of exotic final states. 
The simulation proceeds in three steps: \textit{(i)} the annihilation cross section is computed at parton level with \texttt{MadGraph v5}~\cite{Alwall:2014hca}; \textit{(ii)} the decay, showering and hadronization of the final state particles is performed by \texttt{Pythia v8}~\cite{Sjostrand:2014zea}; \textit{(iii)} the output of the generation is the energy spectrum $dN/dE$ of a given stable particle species.

Examples of the gamma ray spectra produced in different DM annihilation channels are provided in figure \ref{gammas}. The main effect of having exotic particles in the final state, compared to annihilation into the SM only, is that the spectrum is shifted to the left due to the larger multiplicity of SM particles in the final state. To obtain indirect detection constraints, \texttt{MadDm v3} includes the Fermi-LAT likelihoods \cite{Fermi-LAT:2016uux} for gamma rays from dwarf spheroidal galaxies. These Milky Way satellite galaxies are the most reliable sources for indirect detection due to their proximity, large DM density and the fact that they are supposedly free from other gamma-ray emission. In contrast, the galactic center (GC), in spite of being the brightest DM source in the sky, has also bright astrophysical backgrounds with large uncertainties. For this reason, we will not consider experiments looking towards the GC, such as H.E.S.S. \cite{Abdallah:2016ygi} which provides more constraining bounds for masses around the TeV. (Moreover, different assumptions on the DM density profile can lead to discrepancies of an order of magnitude in the results of these experiments \cite{Abdallah:2018qtu,Hryczuk:2019nql}.)

The Fermi-LAT collaboration performed a combined analysis for 45 stellar systems among which 28 confirmed DM--dominated dwarfs (Pass 8 data), with a sensitivity to gamma rays with energy in the window between 500 MeV to 500 GeV  \cite{Fermi-LAT:2016uux}. Although specific targets showed some excesses of signal over background, the combined analysis revealed no significant global excess and therefore strong bounds were set on the expected gamma ray flux from DM annihilation:
\begin{equation}
\Phi \left(\Delta \Omega , E_{min}, E_{max}\right) \propto \frac{\left< \sigma v \right>}{m_\eta^2} \int_{E_{min}}^{E_{max}} \frac{d N_\gamma}{dE_\gamma} dE_\gamma \times \int_{\Delta \Omega} \int_{l.o.s.} \rho_{\rm DM}^2 (\vec{r}(l)) dl d\Omega~,
\end{equation}
where $\left< \sigma v \right>$ is the velocity averaged DM annihilation cross section and $N_\gamma$ ($E_\gamma$) is the number (kinetic energy) of prompt photons in the final state. The second term in this equation is the so-called \textit{J}-factor, a geometric function that depends on the integration of the DM density over the solid angle ($\Delta \Omega$) and the line of sight (l.o.s.). These astrophysical factors are provided by the collaboration and adopted from Ref. \cite{Geringer-Sameth:2014yza}.
\begin{figure}[t]
 \begin{center}
  \includegraphics[width=0.496\columnwidth]{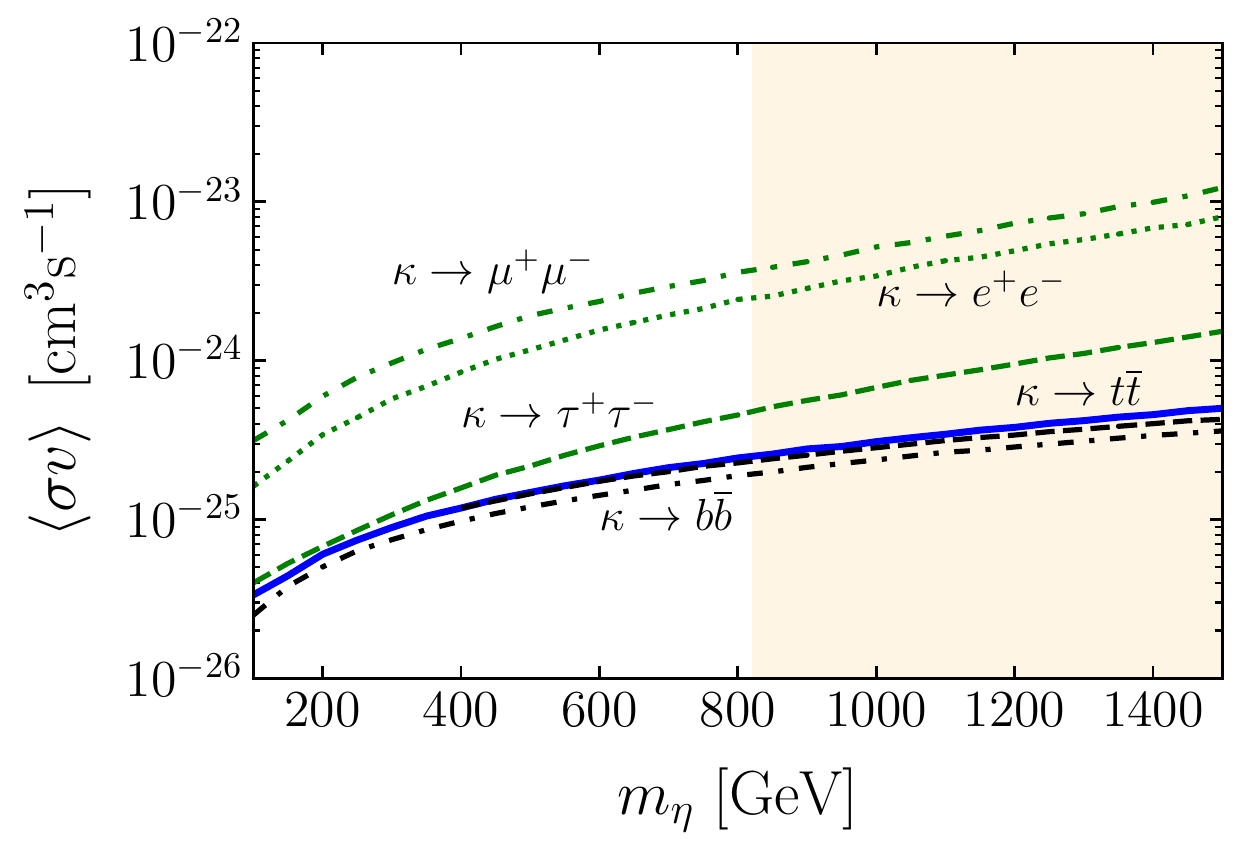}
  \includegraphics[width=0.496\columnwidth]{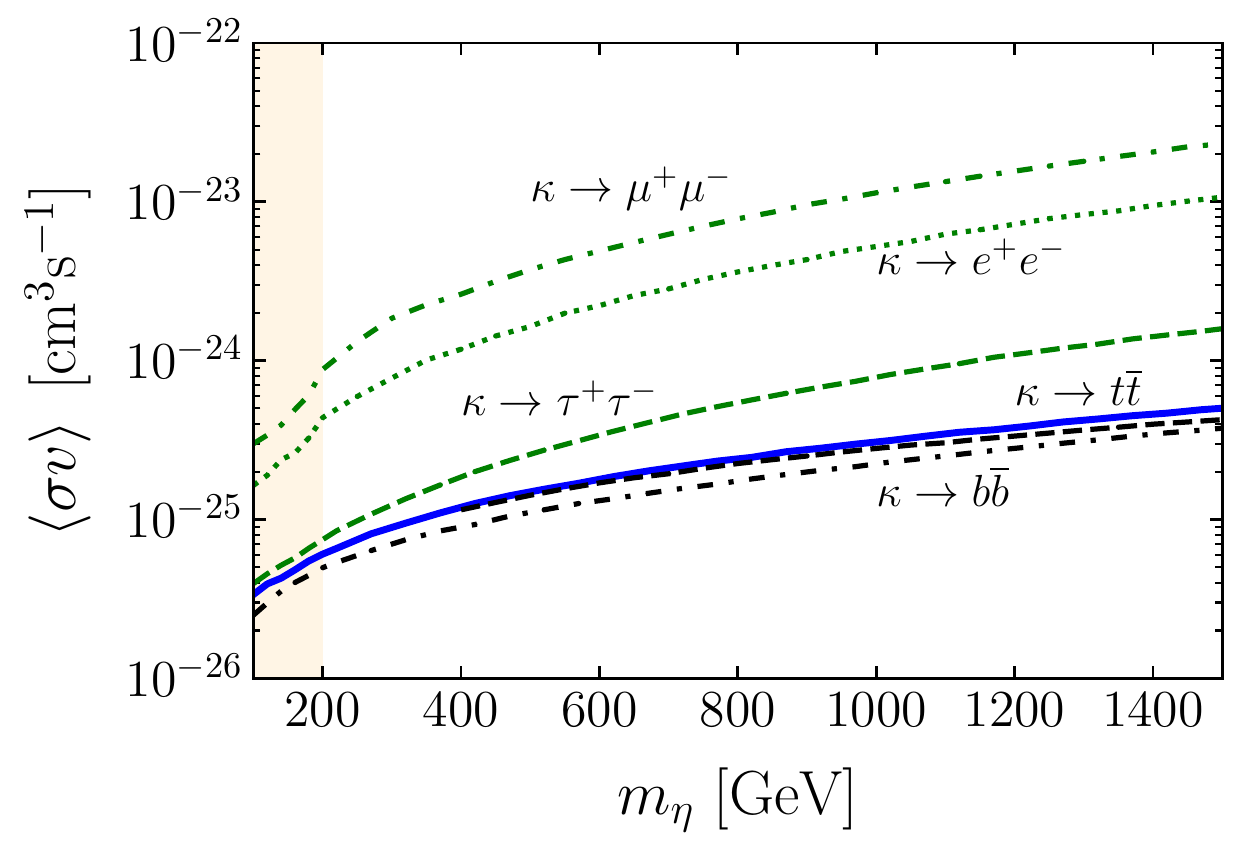}
 \end{center}
\caption{\it Fermi-LAT exclusion limits in the plane $(m_\eta, \left< \sigma v\right>)$. The solid blue line stands for $\eta \eta \rightarrow hh + WW + ZZ$, corresponding to the SM annihilation channel in the limit of large DM masses. In the orange area, $m_\kappa$ is a function of the DM mass to fit $\Omega_{\rm obs} h^2$. The dashed and dotted curves correspond to the annihilation into the new scalar, each for a specific SM final state. The results for RegI and RegII ($f = 2.5$ TeV) are shown in the left and right panels, respectively.}
 \label{id_Fermi}
\end{figure}
From this formula, we are able to confront the particle physics model with the gamma ray flux observed by the experiment, in order to set bounds on the cross section for a given DM mass. 
The results of the simulation are plotted in figure \ref{id_Fermi}, for the different regimes. We have set $m_\kappa = m_\eta$ and $m_\kappa = 20$ GeV outside the orange region, for RegI and RegII ($f=2.5$ GeV), respectively. 

For comparison, we consider the bosonic SM channel, since the DM annihilation to fermions is suppressed by $m_f^2/ ({\rm few}~m_\eta^2)$, where $m_f$ is the fermion mass. As sustained in previous sections, a large fraction of DM annihilates into $\kappa\kappa$. In this case, the Fermi bounds for leptonic decays (especially final states with electrons and muons) are much weaker than those for the hadronically decaying scalar and the SM $hh, WW, ZZ$ channels. This can be explained because quark hadronization produces a large number of neutral pions, which then decay into low-energy gamma rays (with a branching ratio of $\sim 99\%$) \cite{PhysRevD.98.030001} making the gamma spectrum much larger.

Although we do not show it explicitly, we remark that the new Fermi bounds are at most an order of magnitude apart from those corresponding to the annihilation into the SM only, \textit{i.e.} $\eta\eta\to\kappa\kappa,\kappa\to\psi\overline{\psi}$ versus $\eta\eta\to \psi\overline{\psi}$.
The main discrepancy arises in leptonic channels, mainly muons, whose photon spectrum originates mainly from final state radiation and becomes very flat; see figure \ref{gammas}. Therefore, the inclusion of an intermediate step smooths out the spectrum leading to significant modifications. On one side, the fact that the spectrum peaks at lower  $x = E_\gamma /m_\eta$ leads to a larger number of low-energy photons. Some of these photons may lie outside the Fermi window or in low-energies bins, where the backgrounds are generally larger, weakening the bounds. 
On the other side, for large DM masses, the intermediate step might rather strengthen the constraints by moving the spectrum that is above the $500$ GeV threshold back to the Fermi window \cite{Elor:2015bho}.

If a thermal $\eta$ explains the totality of the DM abundance, it must have an annihilation rate of the order of  $\left< \sigma v \right> \sim 3 \times 10^{-26}\rm{cm}^{3}~\rm{s}^{-1}$. This cross section is unconstrained by the Fermi data in both regimes. Next-generation Cherenkov telescopes, with an improved capability to discriminate the GC radiation components, such as CTA \cite{Acharya:2017ttl} might strengthen the Fermi bound by one-to-two orders of magnitude \cite{Hryczuk:2019nql}. (Future wide field-of-view gamma-ray observatories might provide competitive bounds\cite{Viana:2019ucn}; interestingly, they can be less sensitive to the DM profile in the GC.)
In this case, both the SM and $\kappa \rightarrow q \overline{q}$ channels could be excluded under the first assumption, while $\kappa \rightarrow \ell^+\ell^-$ ($\ell = e, \mu$) remains viable.
Current constraints from the AMS-02 \cite{Aguilar:2016kjl} antiproton spectra are potentially powerful for final states with electrons and muons; however, the size of systematic uncertainties is still debatable \cite{Balkin:2018tma}. 
A conservative analysis including these bounds, in the context of DM annihilating directly to SM particles, show that indeed muons are the least constrained among all visible annihilation products \cite{Leane:2018kjk}. Moreover, the corresponding bounds are less sensitive to the mass of the  intermediate state than other leptonic bounds \cite{Elor:2015bho}.

We finally remark that, while within our CHM there is generally no freedom for $c_\eta$ in equation \ref{eq:Ly}, 
$c_\kappa$ can be chosen independently. 
In general, quarks and leptons can have different embeddings, with $\gamma_q \neq \gamma_\ell$ in equation \ref{eq:Yuk27}, because they couple to different operators of the composite sector.  The exotic scalar can therefore be naturally leptophilic provided that $\gamma_q = 0$ in this equation (\textit{i.e.} provided that the quark sector respects a $\kappa\rightarrow -\kappa$ symmetry); see Ref. \cite{Blance:2019ixw}  for the case of a muonphilic singlet. The challenge for probing this scenario with indirect detection searches would be extremely hard. Other searches, such as collider analyses, are instead mandatory.

\begin{figure}[t]
 \includegraphics[width=0.49\columnwidth]{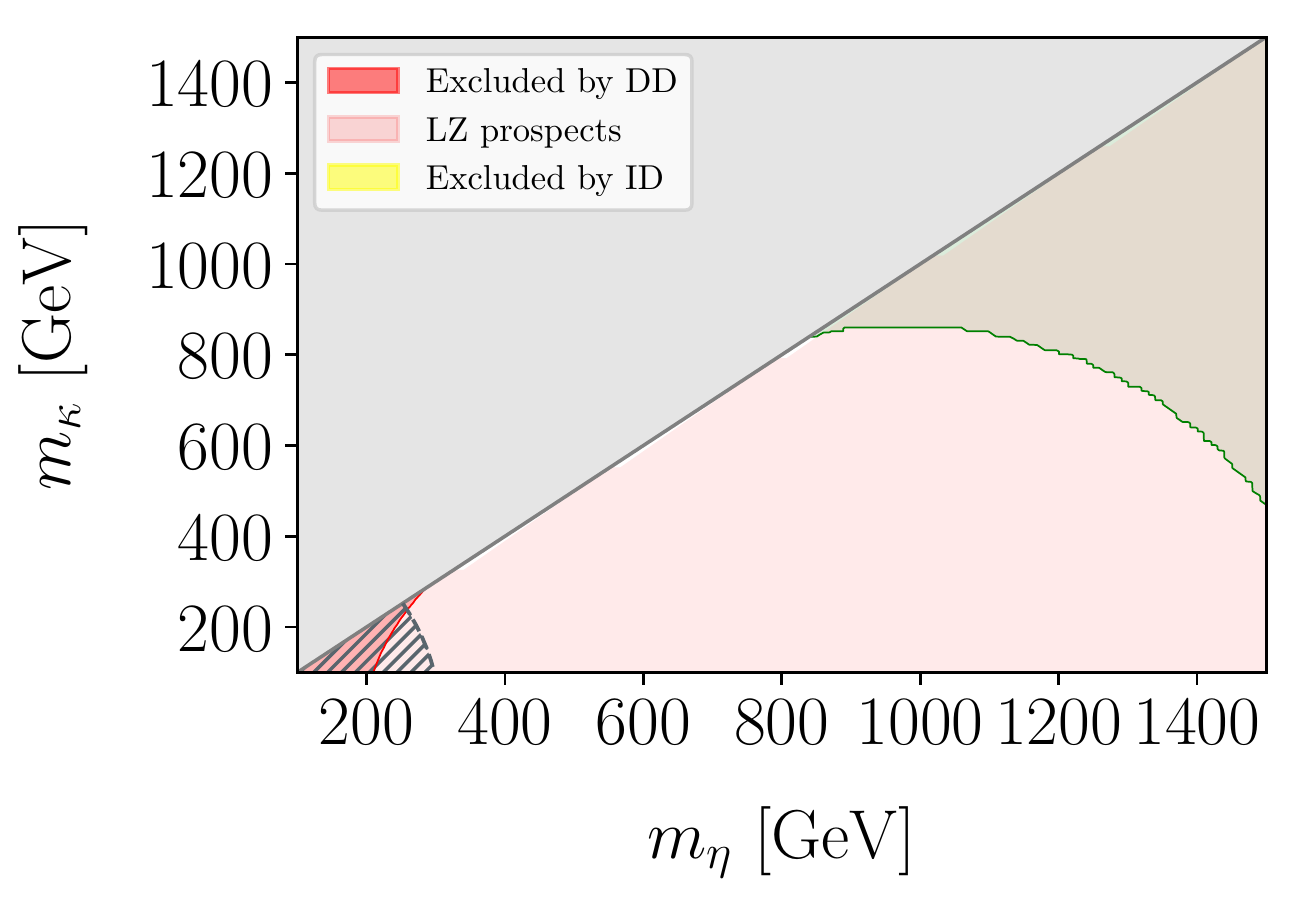}
 \includegraphics[width=0.49\columnwidth]{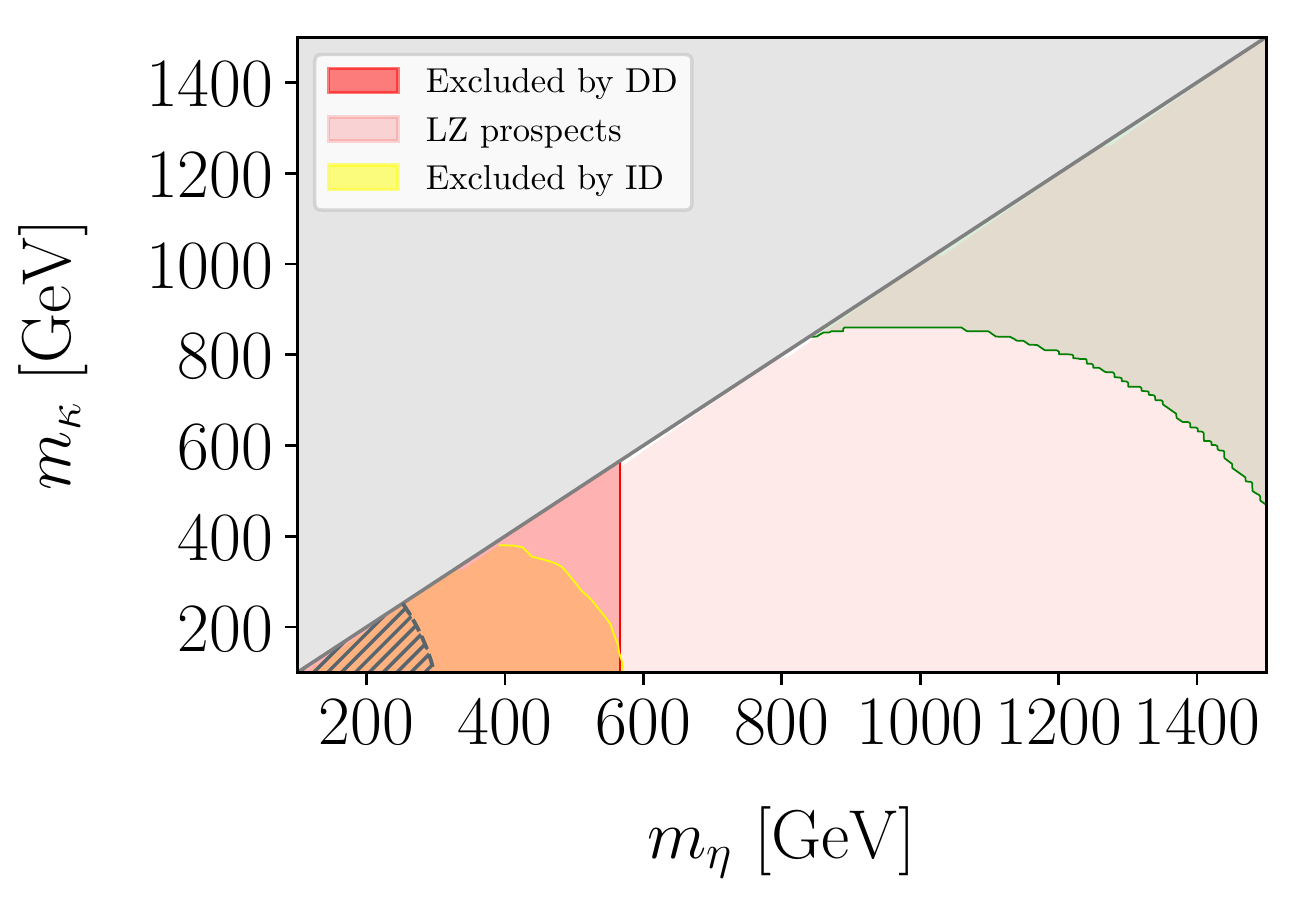}
\includegraphics[width=0.49\columnwidth]{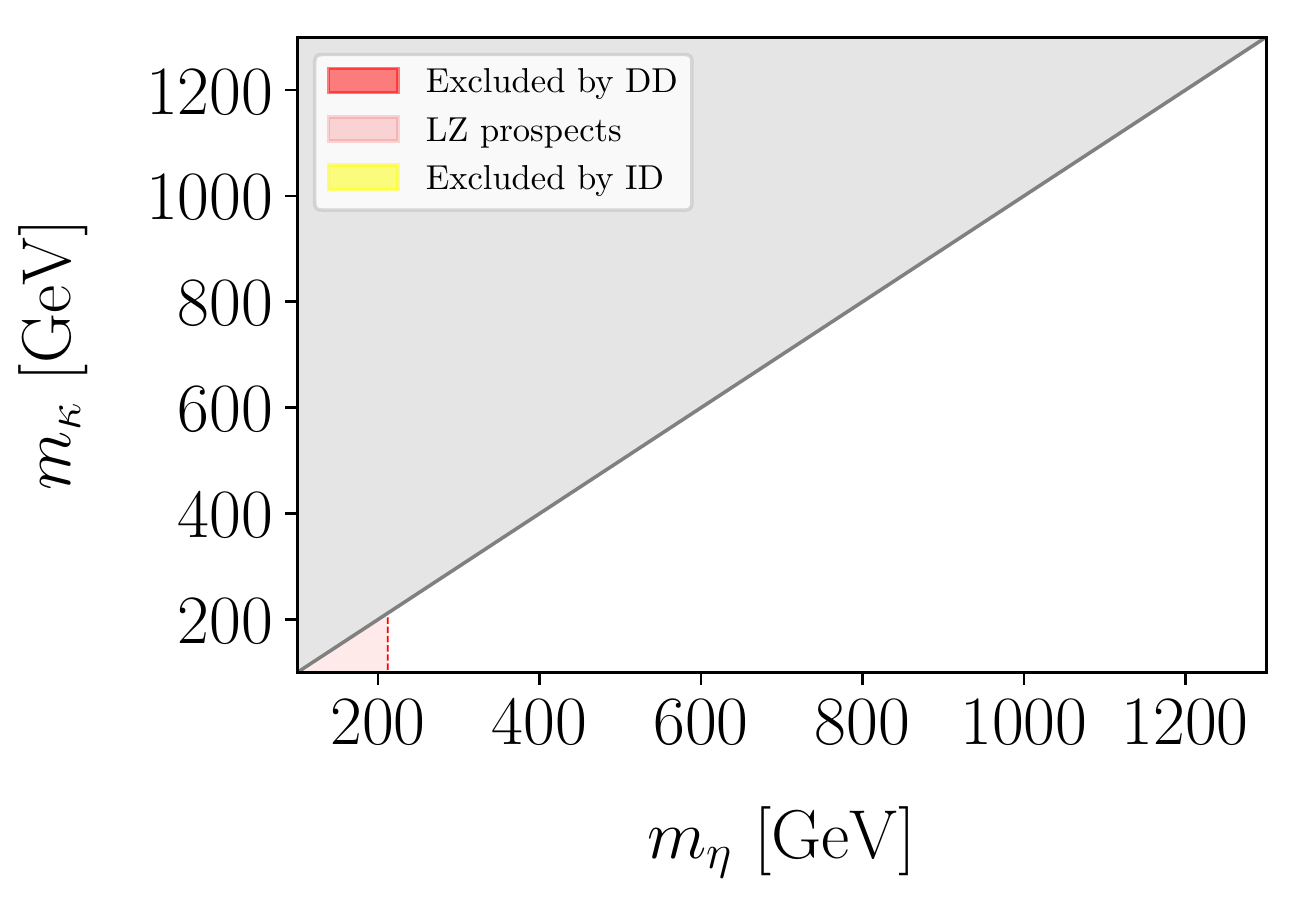}
 \includegraphics[width=0.49\columnwidth]{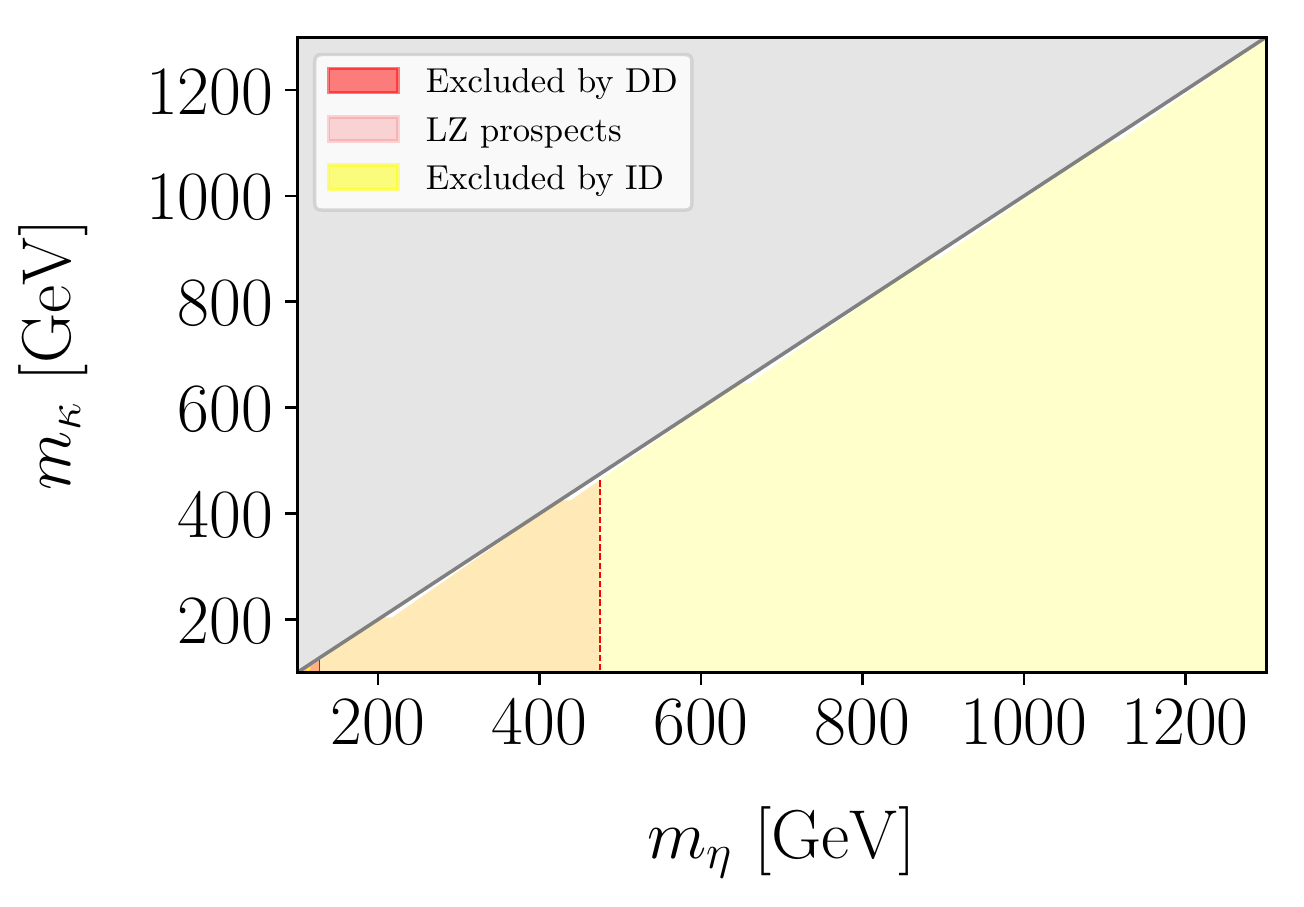}
 \caption{\it Combination of different DM constraints, for $c_\eta = c_\kappa = 1$. The green region is excluded by the relic density. The red and yellow regions are excluded by direct and indirect detection, respectively; finally, the red light region is expected to be probed by LZ.  In the slashed region, $f$ is excluded by EWPD.
 The upper left (right) panel corresponds to the thermal (non-thermal) scenario in RegI. The same holds for the bottom left and right panels in RegII, for $f = 1$ TeV.}
\label{fig:allRegI}
\end{figure}

Assuming instead that $\kappa$ couples to all fermions, we plot in figure \ref{fig:allRegI} all the DM constraints for each regime.
The green region is excluded by the relic density; while the red and yellow regions are excluded by the current direct and indirect searches, respectively. The projected sensitivity for the LZ experiment excludes the light red region. The panels on the left concern the case of thermal DM: we assume that $\eta$ is produced solely via the freeze-out mechanism within a standard cosmology. In the case that $\eta$ is under-abundant, we account for the possibility that the missing fraction of DM originates from other species. Therefore, the fluxes for direct and indirect detection are rescaled by $r = \left(\Omega h^2\right)/ \left(\Omega_{\rm obs} h^2\right)$ and $r^2$, respectively. 
In the right panels, on the other hand, we consider that $\eta$ makes up all DM: regardless of the thermal abundance, points $(m_\eta, m_\kappa)$ outside the green region can match the observed relic density if additional contributions from non-thermal production are taken into account; see Ref. \cite{Allahverdi:2012wb} for an example. No rescaling is therefore applied.
As it can be seen, the non-thermal scenario in RegII with $f=1$ TeV is excluded by indirect detection. In contrast, RegII with thermal DM escapes all constraints. (Considering larger new physics scales, the model is further constrained by the relic density; however, wide regions of the parameter space remain viable for $f < 4$ TeV.)

In such regime, neither direct nor indirect DM searches could be sensitive to these models in the near future. Together with the leptophilic scenario, this shows clearly the necessity for collider analyses.
We propose some of these in the next section.

\section{Collider signals}\label{sec:colliders}
The DM and the extra scalar are hard to produce directly in $pp$ collisions, as they are singlets of the gauge group.
They can, however, be produced in the decay of heavier composite resonances. The observation of such decay chains is also
required to establish the composite nature of the model.

Among the fermionic resonances, we find bottom ($B$)
and top ($T$) vector-like partners.
They are of special interest because they are required by partial compositeness, ${L \sim \lambda_{t_R} f \overline{q_L} \mathcal{O}_R^t + \lambda_{t_L} f \overline{t_R} \mathcal{O}_L^t + \lambda_{b_R} f \overline{q_L} \mathcal{O}_{R}^b + \lambda_{b_L} f \overline{b_R} \mathcal{O}_L^b + \rm{h.c.}}$ \cite{Serra:2015xfa}.  We have included two different operators mixing with the left-handed quarks, as it is in general required to generate both the top and bottom yukawa couplings. Moreover, we consider an extended global symmetry, $SO(7) \times U(1)_X$, where the latter is assumed to be unbroken. This is necessary to obtain the correct quantum numbers of the SM quarks. As usual, the hypercharge is defined along the third $SU(2)_R$ generator $T_R^3 \equiv \left( T^{12} - T^{34}\right)/\sqrt{2}$ and the newly introduced $U(1)_X$ generator, $Y = T_R^3 + X$.

The bottom and top partners can be arranged in multiplets of the unbroken group $SO(6)$; the lightest of which having a
mass $M\sim f$. In general, this can be either a singlet or a sextuplet.
Let us focus on the case where $q_L$ and $t_R$ are embedded, respectively, in the $\mathbf{27}$ and $\mathbf{1}$ representations (the lightest VLQ could be also the $\mathbf{20}$ in the decomposition $\mathbf{27} = \mathbf{1} \oplus \mathbf{6} \oplus \mathbf{20}$, but we will disregard this case). 

The relevant phenomenology of the top partner singlet of $SO(6)$, $T^\mathbf{1}$, is determined from the linear coupling: 
\begin{align}\label{eq:vlqdecay1}
 L_\mathbf{1} & = \cdots +f U^i_7 U^j_7 \left(\overline{Q_L^{\mathbf{27}}}\right)_{ij} T^\mathbf{1}_R~ +\text{h.c.} \nonumber\\
& = \left[... - h \overline{t_L} - i\gamma\frac{v}{f}\kappa \overline{t_L} \right]T_R + \text{h.c.}~.
\end{align}
Analogously, the relevant couplings of the top partner in the sexplet, $T^\mathbf{\mathbf{6}}$, read:
\begin{align}\label{eq:vlqdecay6}
 L_\mathbf{6} & = \cdots +f U^i_7 U^j_k \left(\overline{Q_L^{\mathbf{27}}}\right)_{ij} \left(T^\mathbf{6}_R\right)^k ~ +\text{h.c.} \nonumber\\
 & =\frac{3}{4}\frac{v}{f} \eta \overline{t_L} T'_R +  \left[\frac{3}{4}\frac{ v}{ f}\kappa \overline{t_L} - i\gamma\frac{1}{2} h \overline{t_L}  \right] T''_R + \text{h.c.}~,
\end{align}
where $T'$ and $T''$ are the two singlets in the decomposition of the $\mathbf{6} = \mathbf{4} \oplus \mathbf{1} \oplus \mathbf{1}$ under $SO(4)$. While the top sector composite operators transform in the $\mathbf{27}_{2/3}$ representation of $SO(7)\times U(1)_X$, the ones coupling to the bottom quark transform in the $\mathbf{27}_{-1/3}$. In the latter case, the embedding of $q_L$ is obtained from equation \ref{eq:QL_t} with the replacements $i b_L \rightarrow i t_L$, $b_L \rightarrow - t_L$, $i t_L \rightarrow - i b_L$ and $t_L \rightarrow b_L$ (leading to opposite $T_R^3$ charge).  Then, equations \ref{eq:vlqdecay1} and $\ref{eq:vlqdecay6}$ hold similarly for the bottom partner.

These VLQs can be produced in pairs, model-independently, via QCD interactions in $pp$ collisions.  (Contributions from heavy vector resonances could enhance this cross section \cite{Chala:2014mma, Azatov:2015xqa, Araque:2015cna, Barducci:2015vyf}, but we will conservatively neglect them.)
The corresponding cross section drops sharply with $M$, and therefore searches for VLQs in this context are
more suited for future colliders than for the LHC~\cite{Chala:2018qdf}. QCD production of $T'$ leads to
$t\overline{t}+E_T^\text{miss}$. This signal is equivalent to that of pair-produced stops decaying to stable neutralinos
in SUSY. The reach of a future 100 TeV collider has been explored in Ref.~\cite{Cohen:2014hxa}. 
By rescaling the results with the VLQ production cross section,  Ref.~\cite{Chala:2018qdf} has set bounds on the plane $(M, m_\eta)$. In particular, for $L = 1~{\rm ab}^{-1}$ and assuming ${\rm BR}\left(T' \rightarrow \eta t \right) = 1$, resonances of mass below $M \sim 9$ TeV and $m_\eta \sim 3$ TeV can be excluded.

\begin{figure}[t]
 \begin{center}
  \hspace{-0.15cm}\includegraphics[width=0.48\columnwidth]{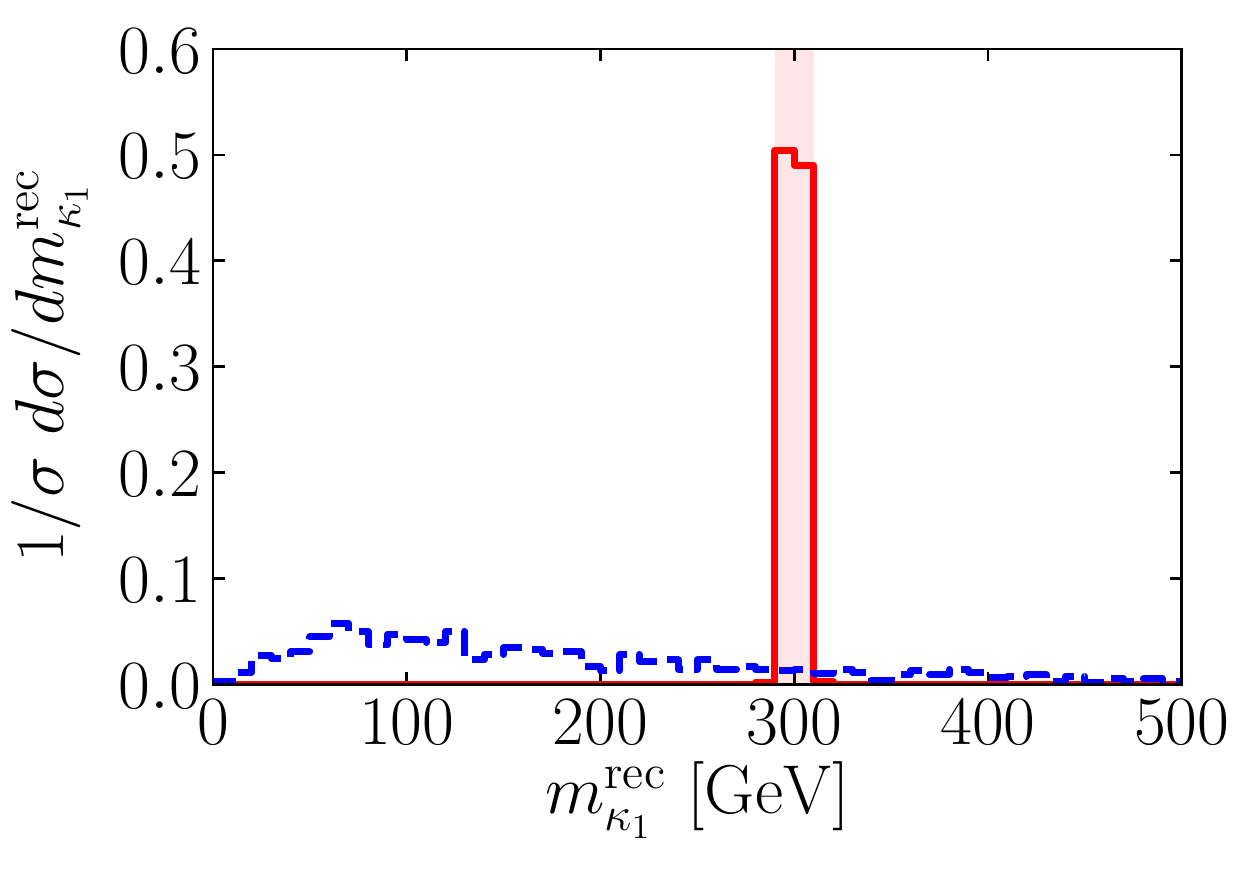}
  \hspace{0.1cm}\includegraphics[width=0.48\columnwidth]{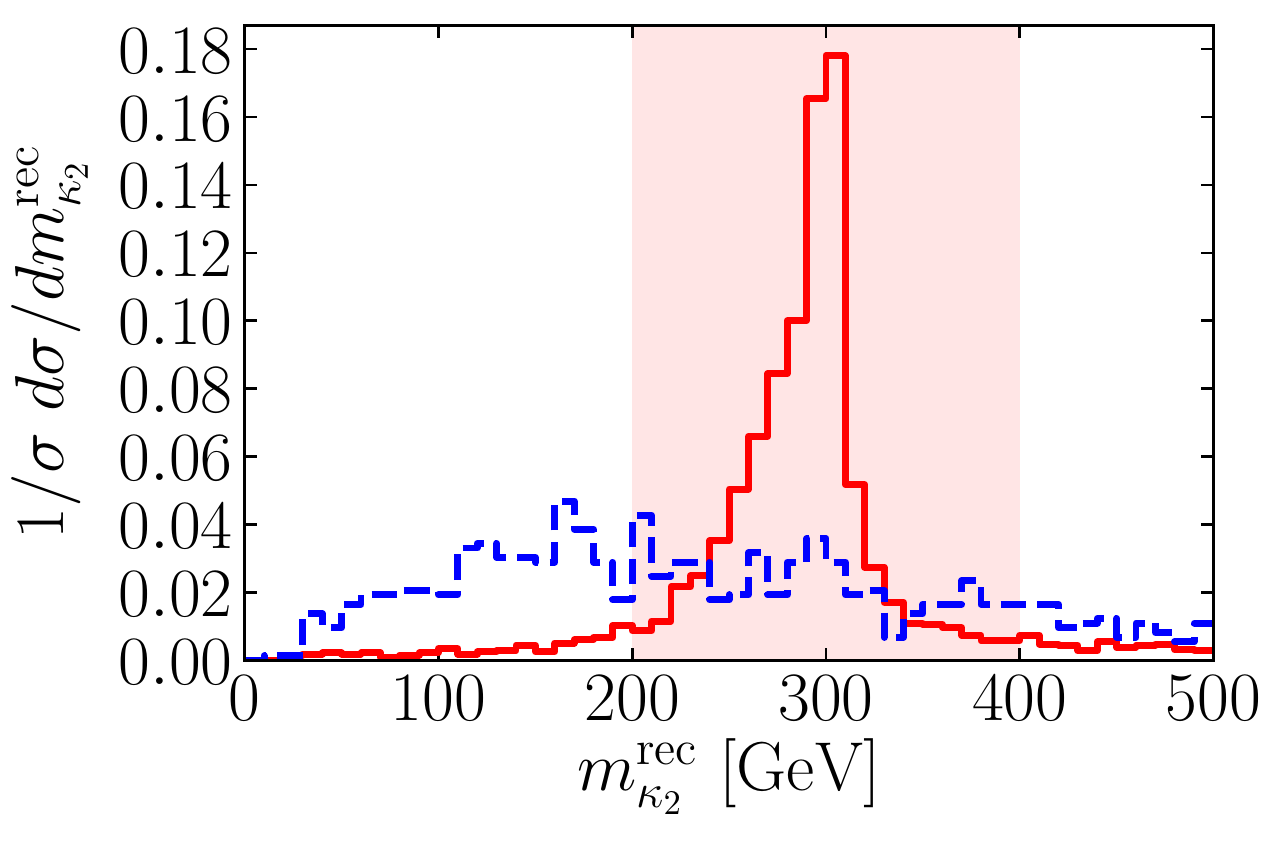}
  \includegraphics[width=0.49\columnwidth]{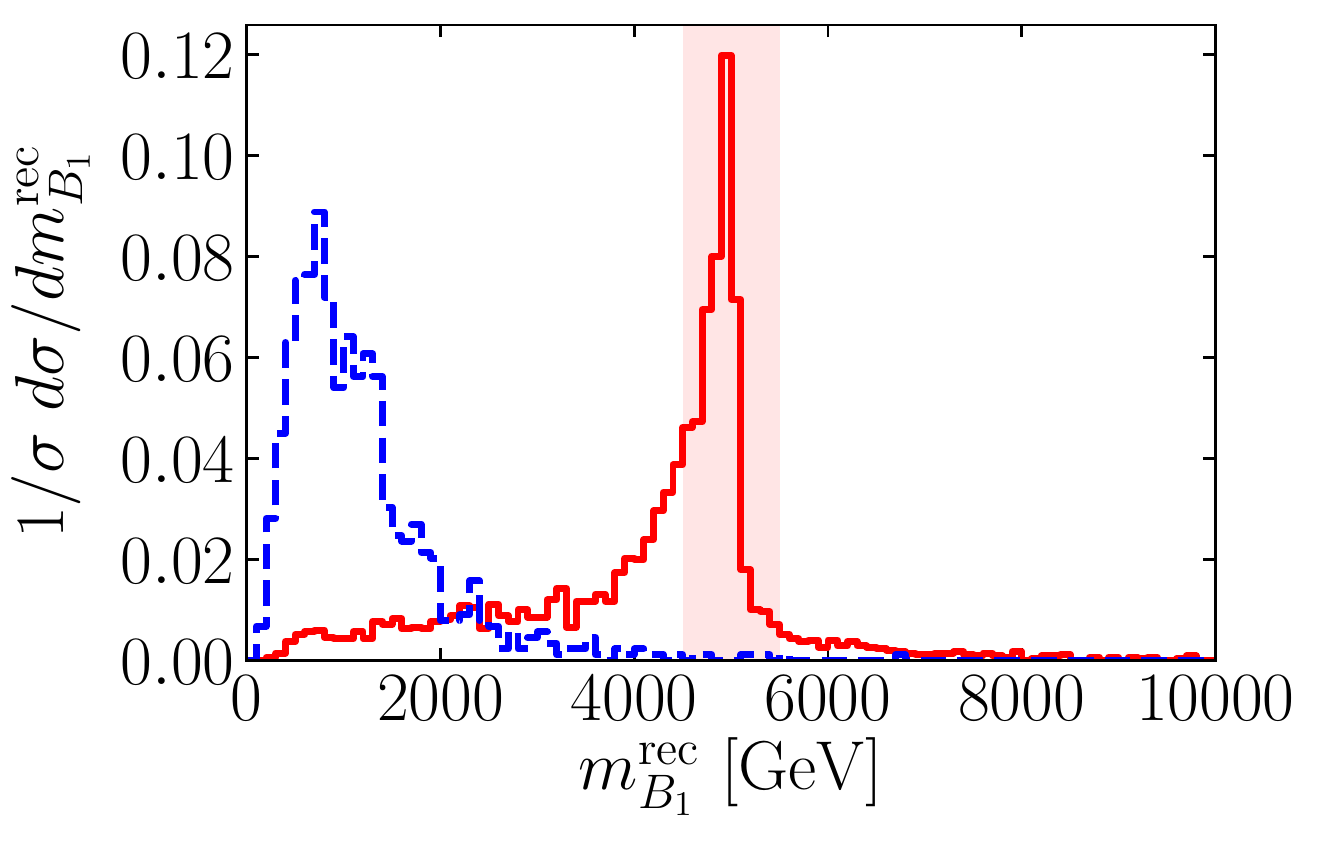}
  \includegraphics[width=0.49\columnwidth]{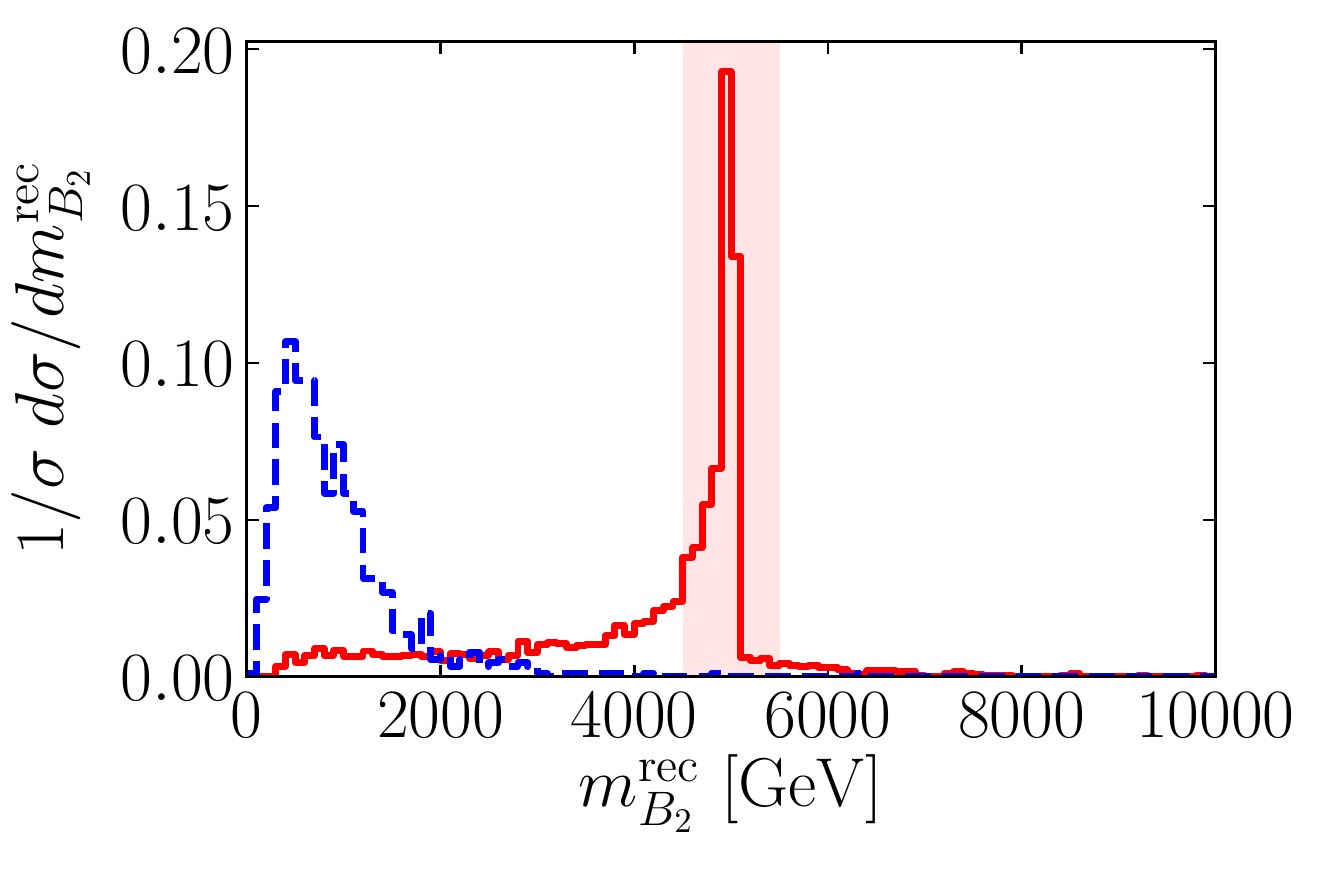}
 \end{center}
 \caption{\it Normalized distributions of $m_{\kappa_1}^\text{rec}$ (upper left), $m_{\kappa_2}^\text{rec}$ (upper right), $m_{B_1}^\text{rec}$ (bottom left) and $m_{B_2}^\text{rec}$ (bottom right) in the signal for $M = 5$ TeV and $m_\kappa = 300$ GeV (solid red) and in the SM background (dashed blue) in the analysis proposed for $pp\to B\overline{B}\to \kappa\kappa b\overline{b}$, $\kappa\kappa\to \gamma\gamma b\overline{b}$. Subsequent cuts are enforced to make the different variables lie within the light red regions.}
 \label{fig:distrVLQa}
\end{figure}
\begin{figure}[t]
 \begin{center}
  \includegraphics[width=0.49\columnwidth]{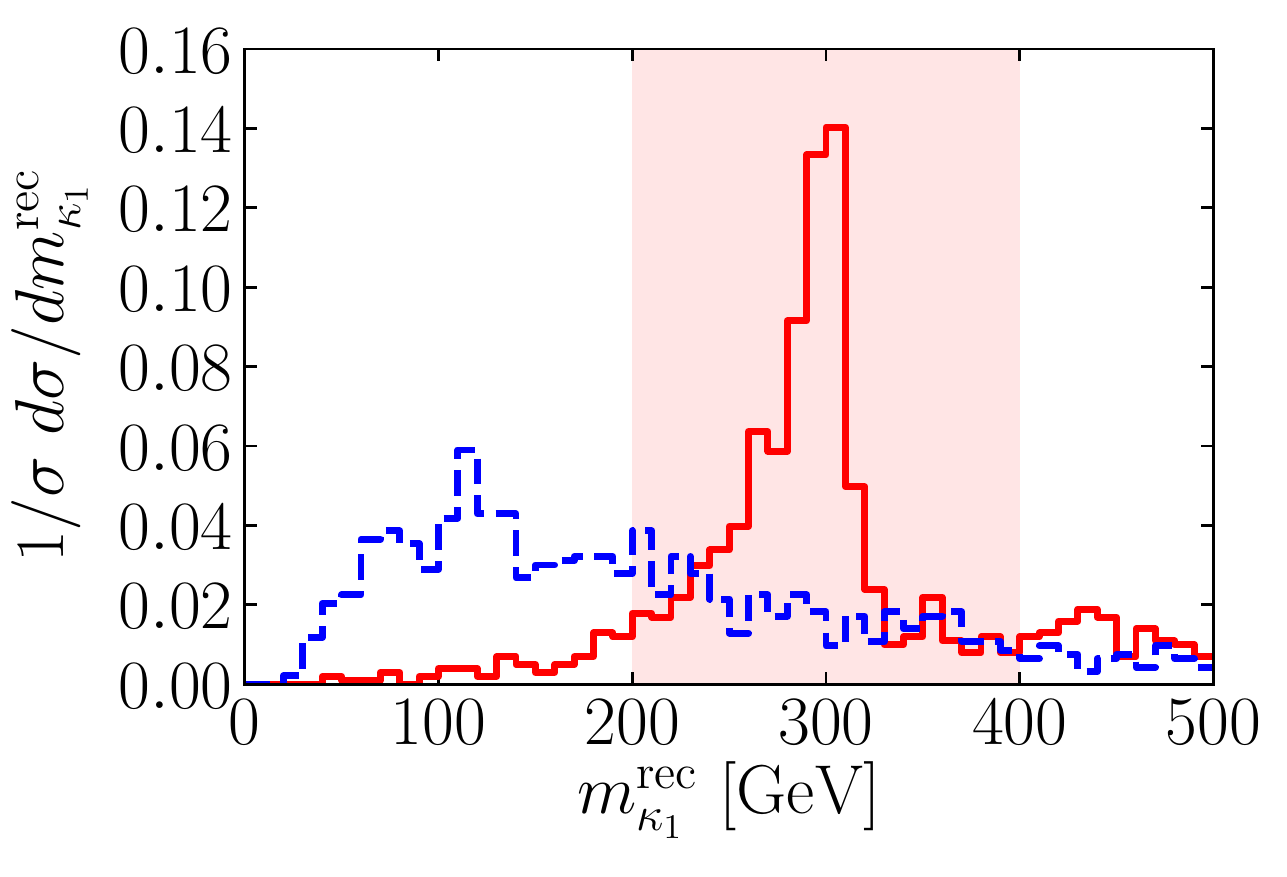}
  \includegraphics[width=0.49\columnwidth]{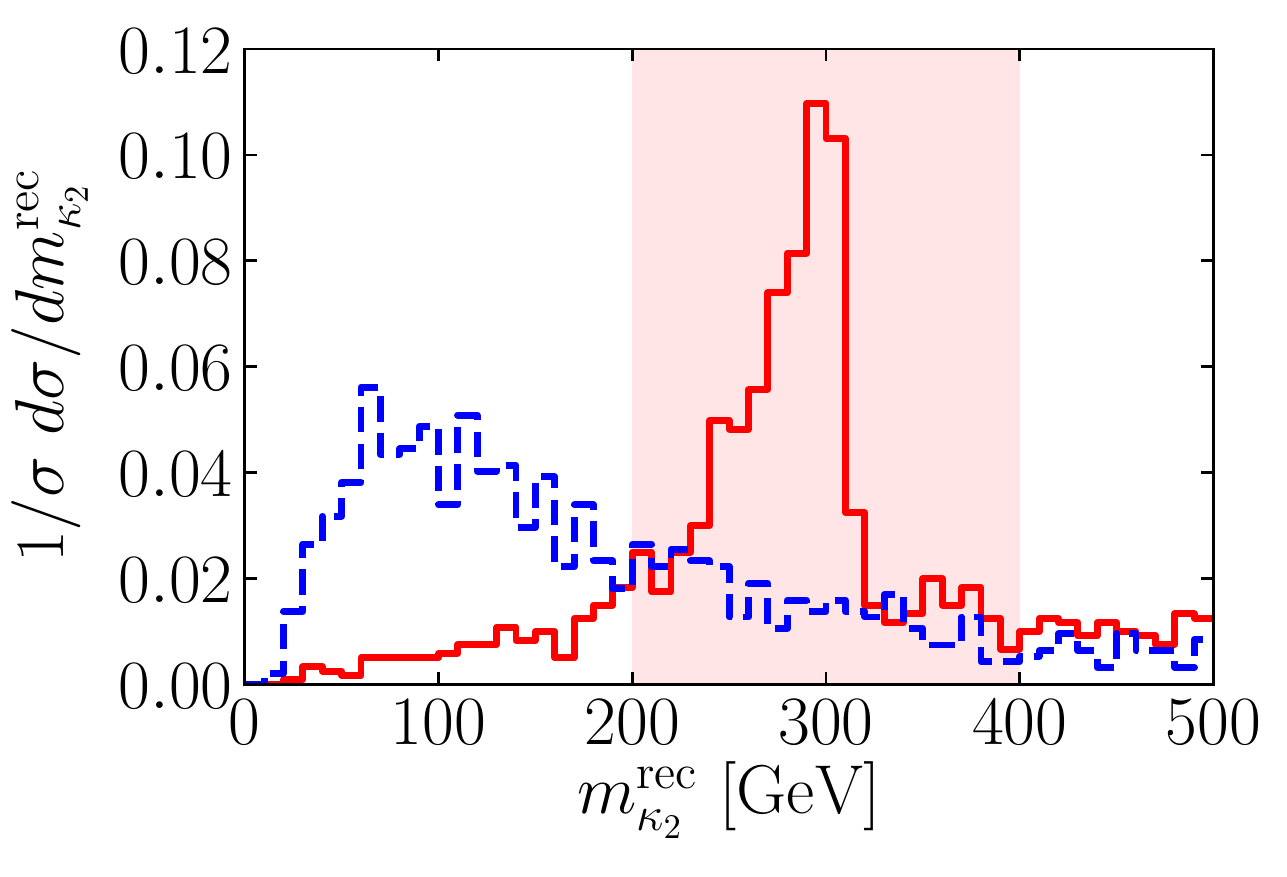}
  \includegraphics[width=0.49\columnwidth]{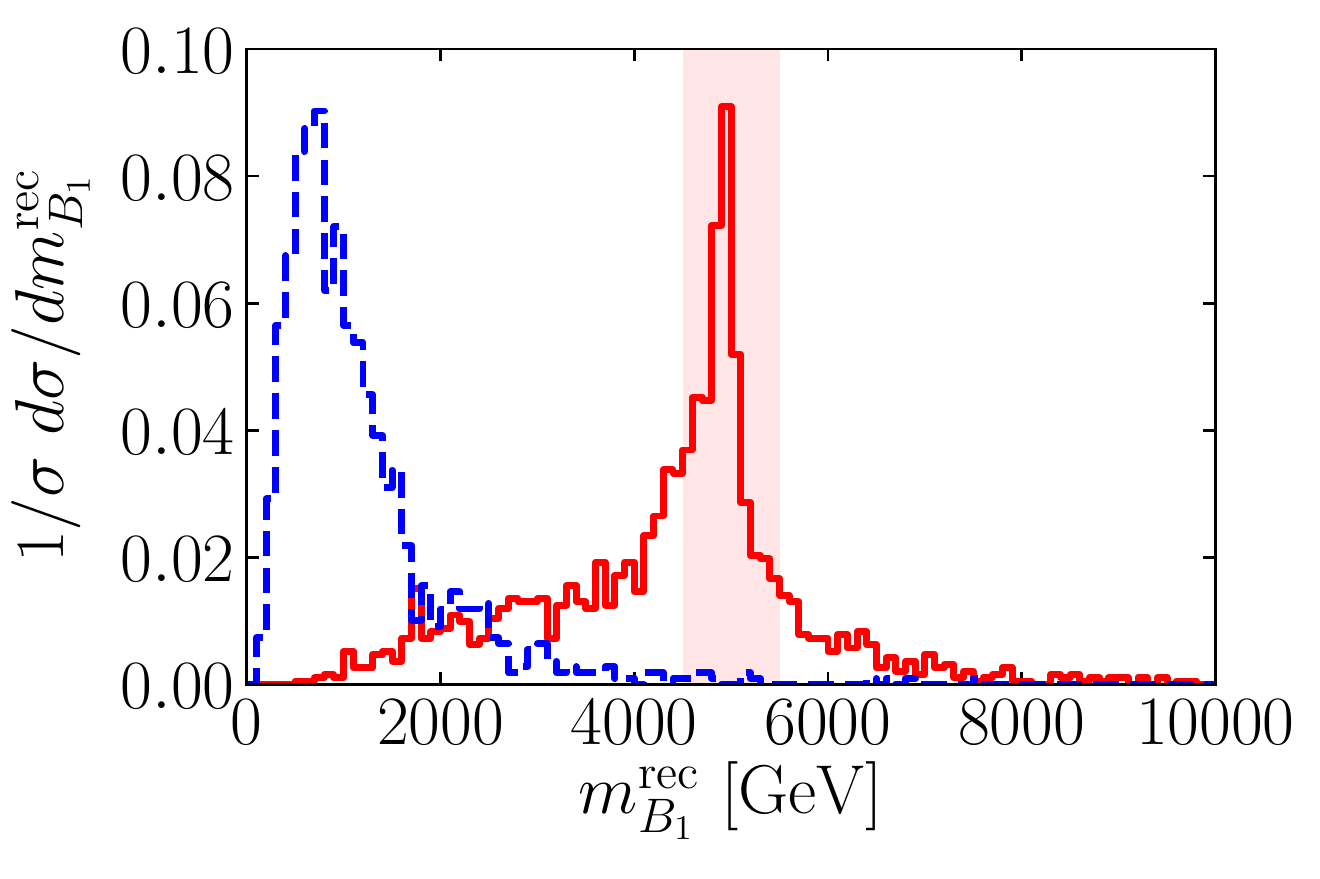}
  \includegraphics[width=0.49\columnwidth]{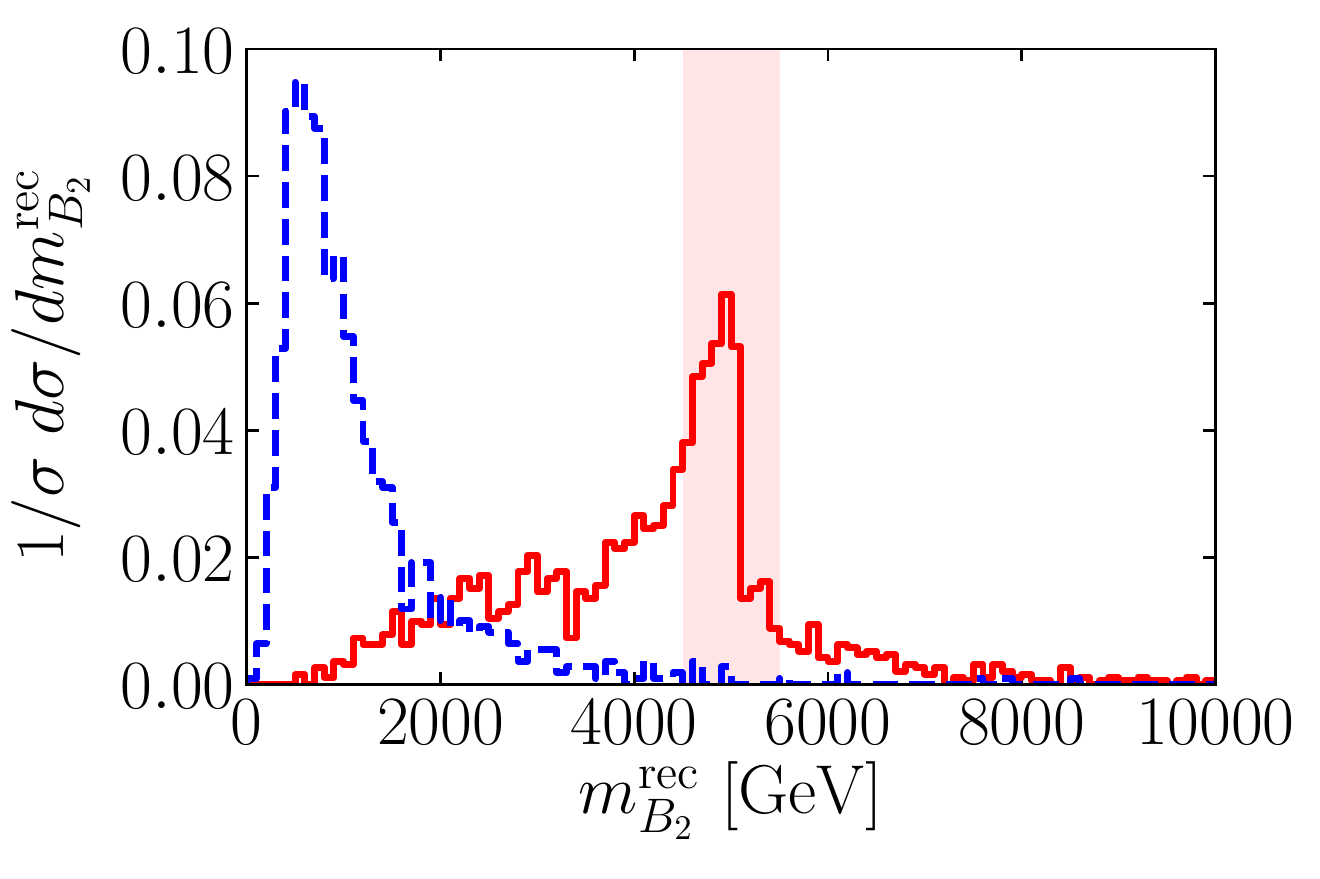}
 \end{center}
 \caption{\it Normalized distributions of $m_{\kappa_1}^\text{rec}$ (upper left), $m_{\kappa_2}^\text{rec}$ (upper right), $m_{B_1}^\text{rec}$ (bottom left) and $m_{B_2}^\text{rec}$ (bottom right) in the signal for $M = 5$ TeV and $m_\kappa = 300$ GeV (solid red) and in the SM background (dashed blue) in the analysis proposed for $pp\to B\overline{B}\to \kappa\kappa b\overline{b}$, $\kappa\kappa\to b\overline{b} b\overline{b}$.}
 \label{fig:distrVLQb}
\end{figure}
\begin{figure}[t]
 \begin{center}
  \includegraphics[width=0.49\columnwidth]{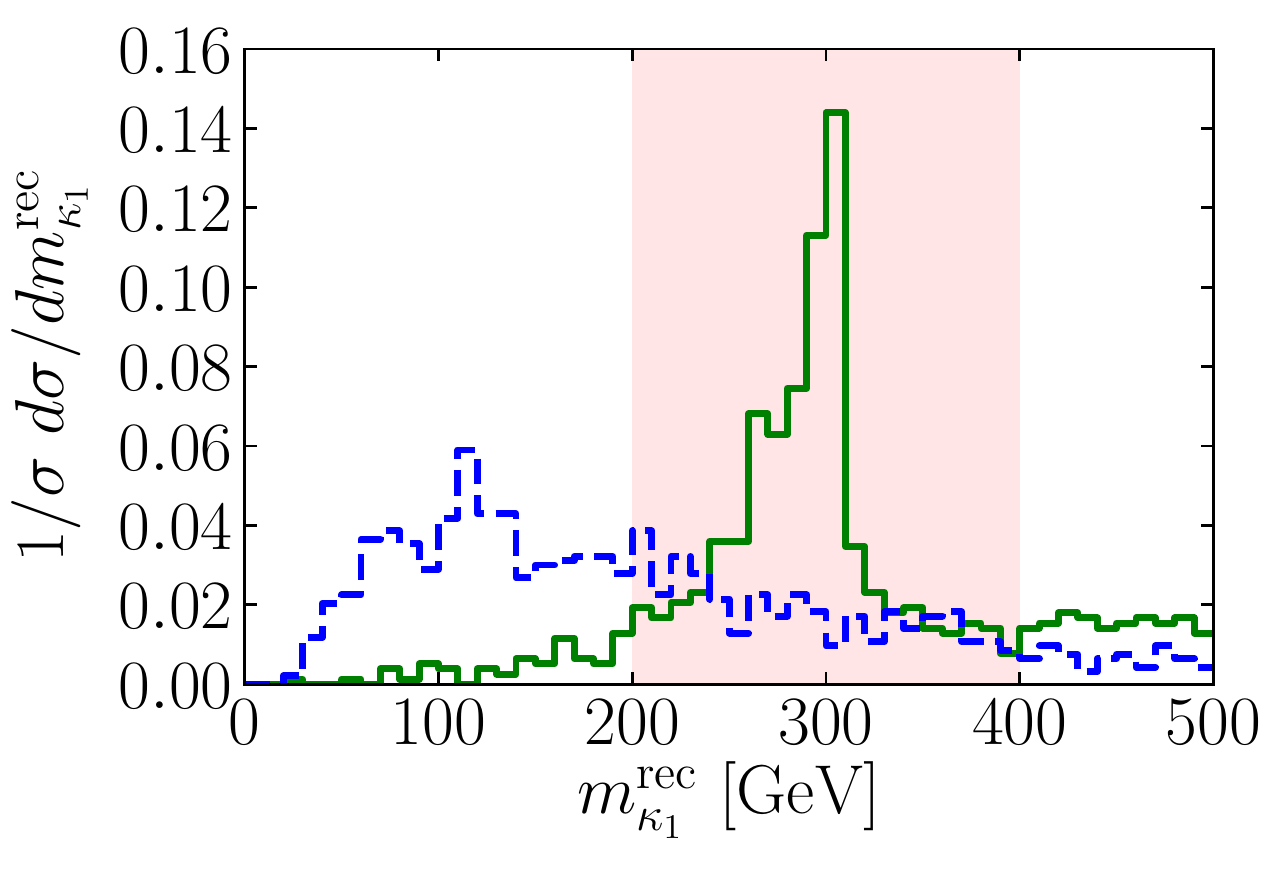}
  \includegraphics[width=0.49\columnwidth]{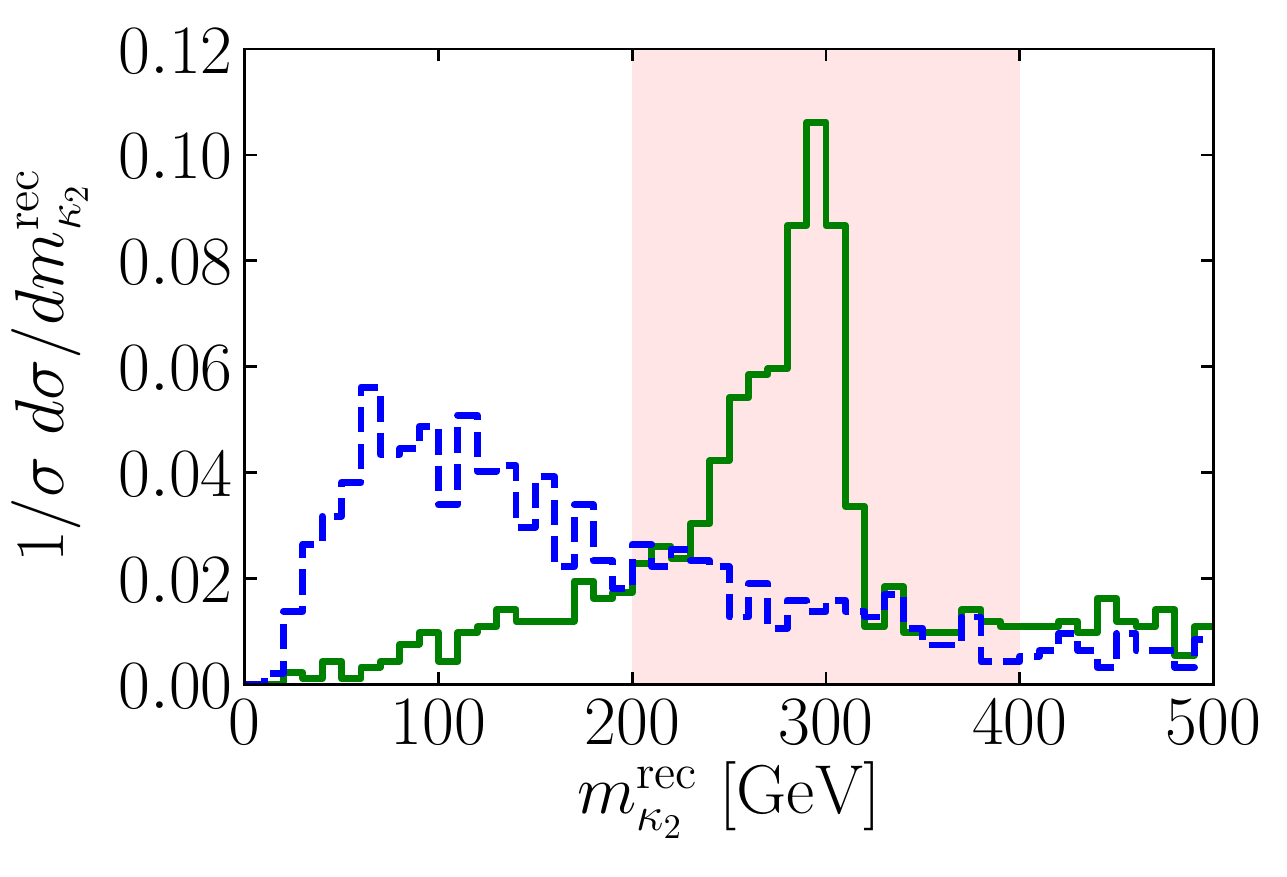}
  \includegraphics[width=0.49\columnwidth]{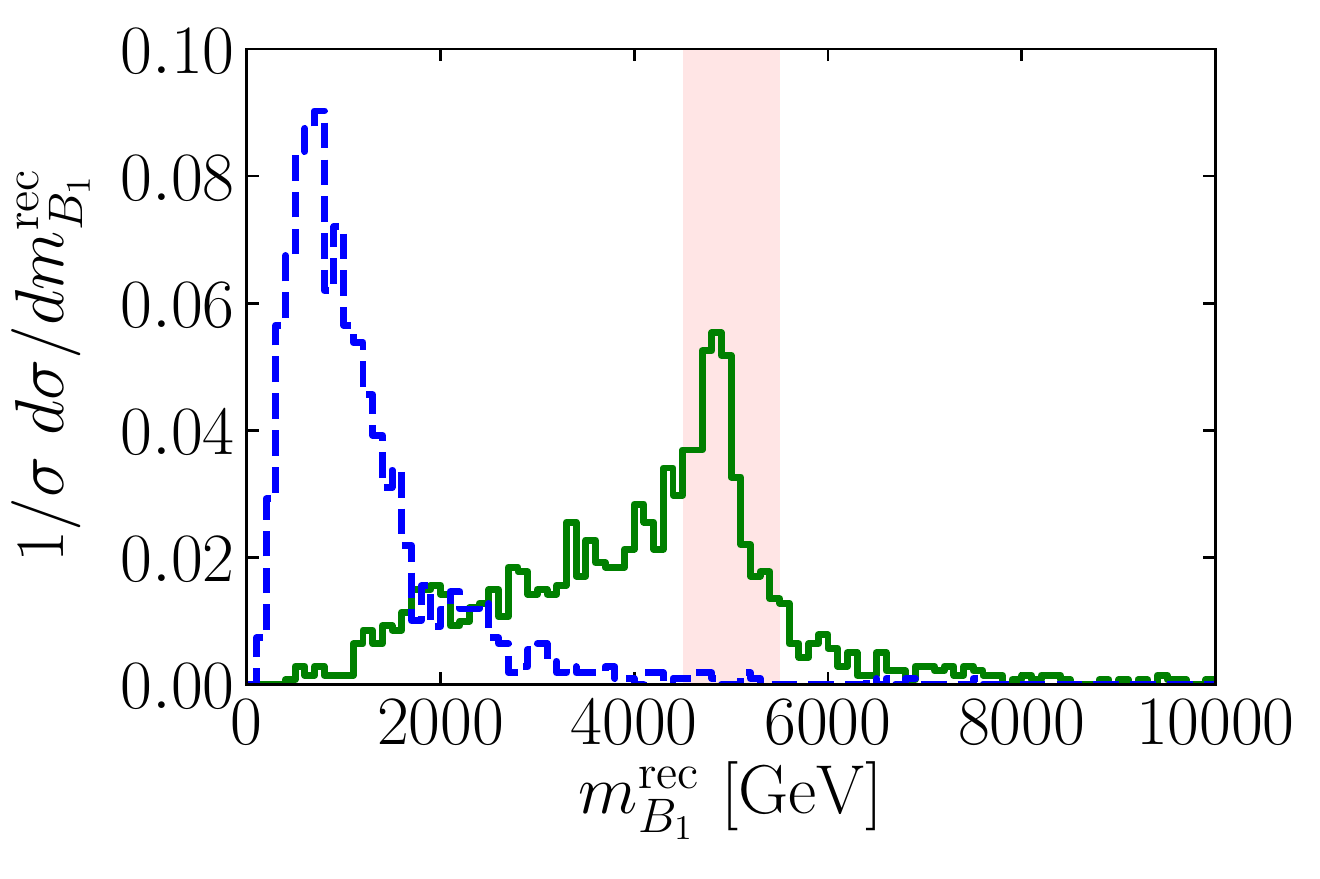}
  \includegraphics[width=0.49\columnwidth]{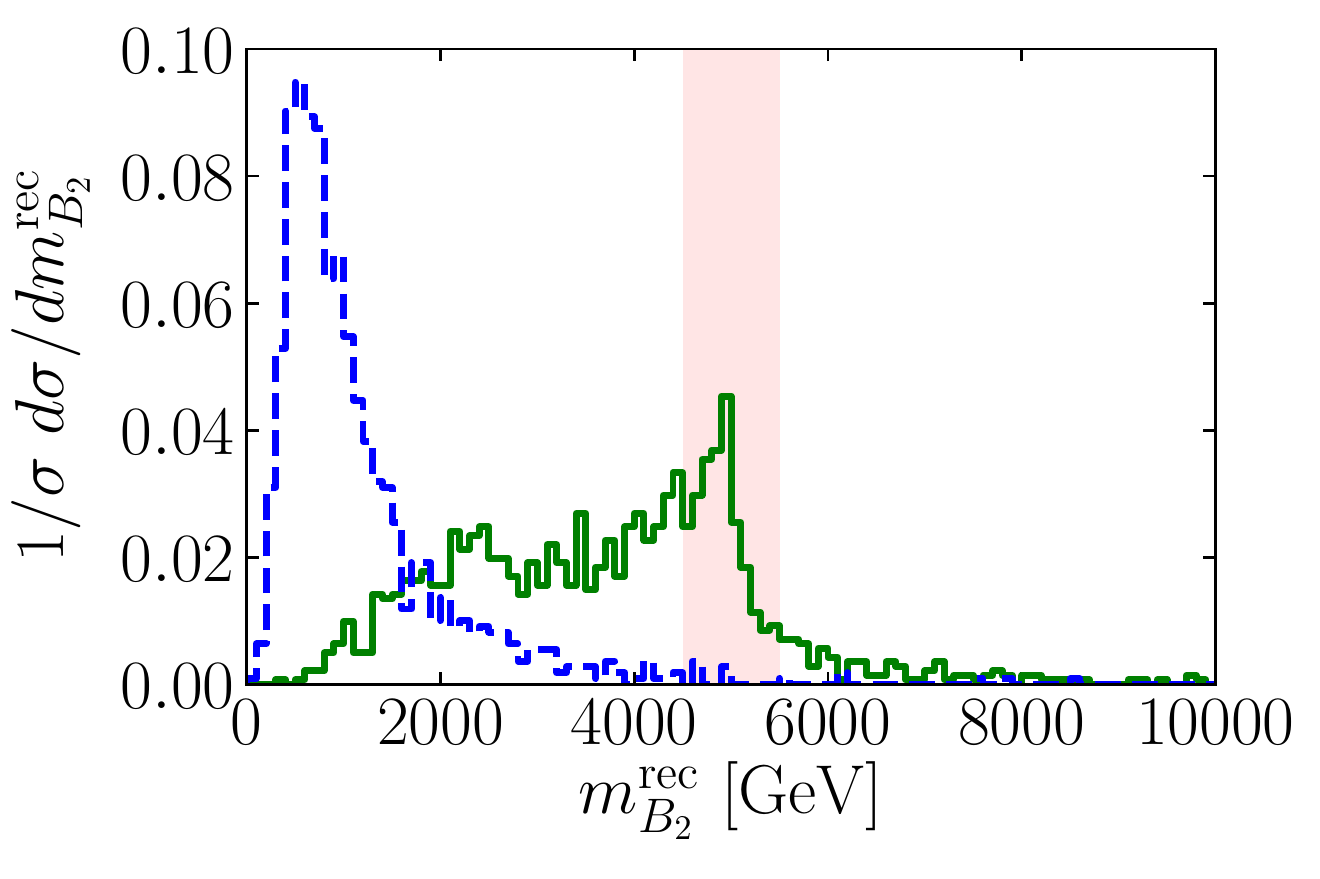}
 \end{center}
 \caption{\it Normalized distributions of $m_{\kappa_1}^\text{rec}$ (upper left), $m_{\kappa_2}^\text{rec}$ (upper right), $m_{B_1}^\text{rec}$ (bottom left) and $m_{B_2}^\text{rec}$ (bottom right) in the signal for a top partner with $M = 5$ TeV and $m_\kappa = 300$ GeV (solid green) and in the SM background (dashed blue) in the analysis proposed for $pp\to B\overline{B}\to \kappa\kappa b\overline{b}$, $\kappa\kappa\to b\overline{b} b\overline{b}$.}
 \label{fig:distrVLQTPb}
\end{figure}
\begin{figure}[t]
 \begin{center}
  \includegraphics[width=0.49\columnwidth]{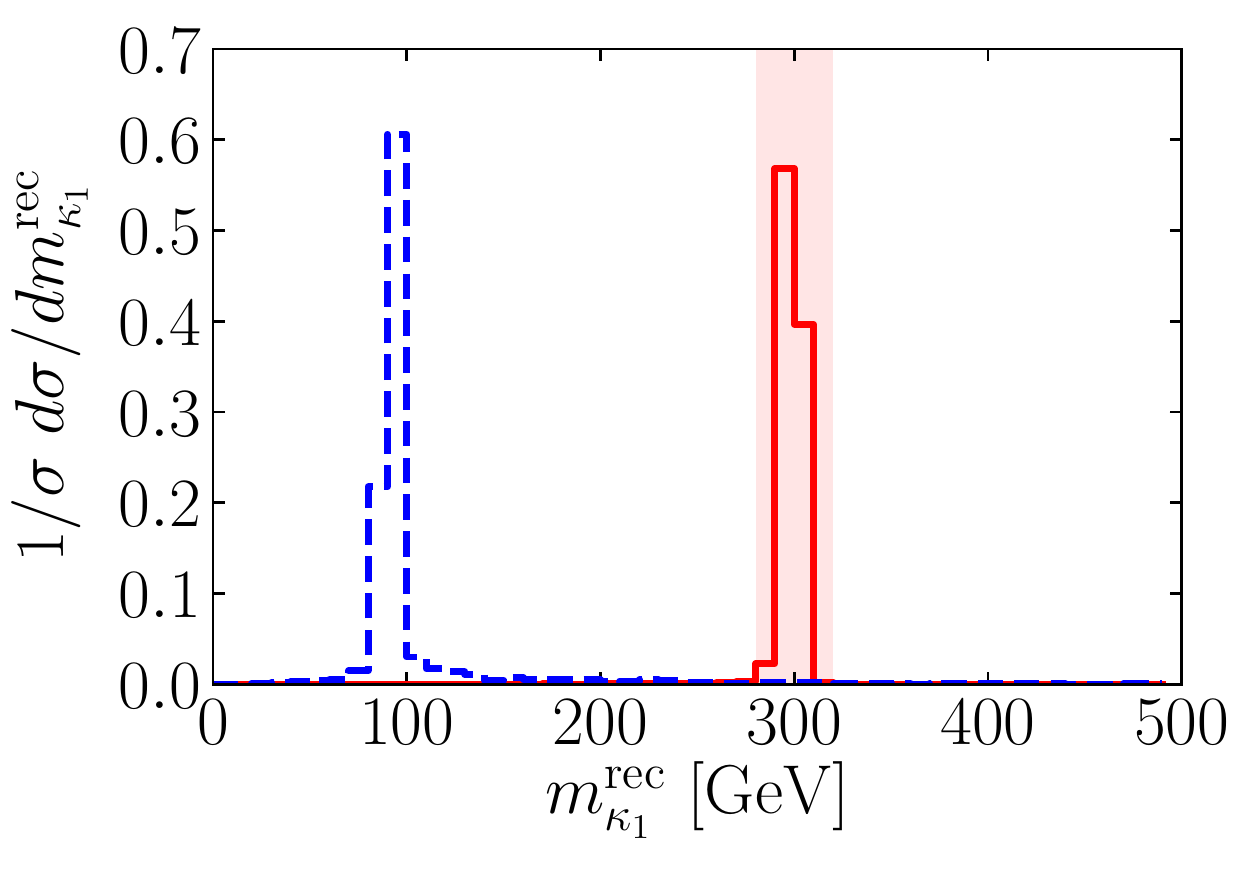}
  \includegraphics[width=0.49\columnwidth]{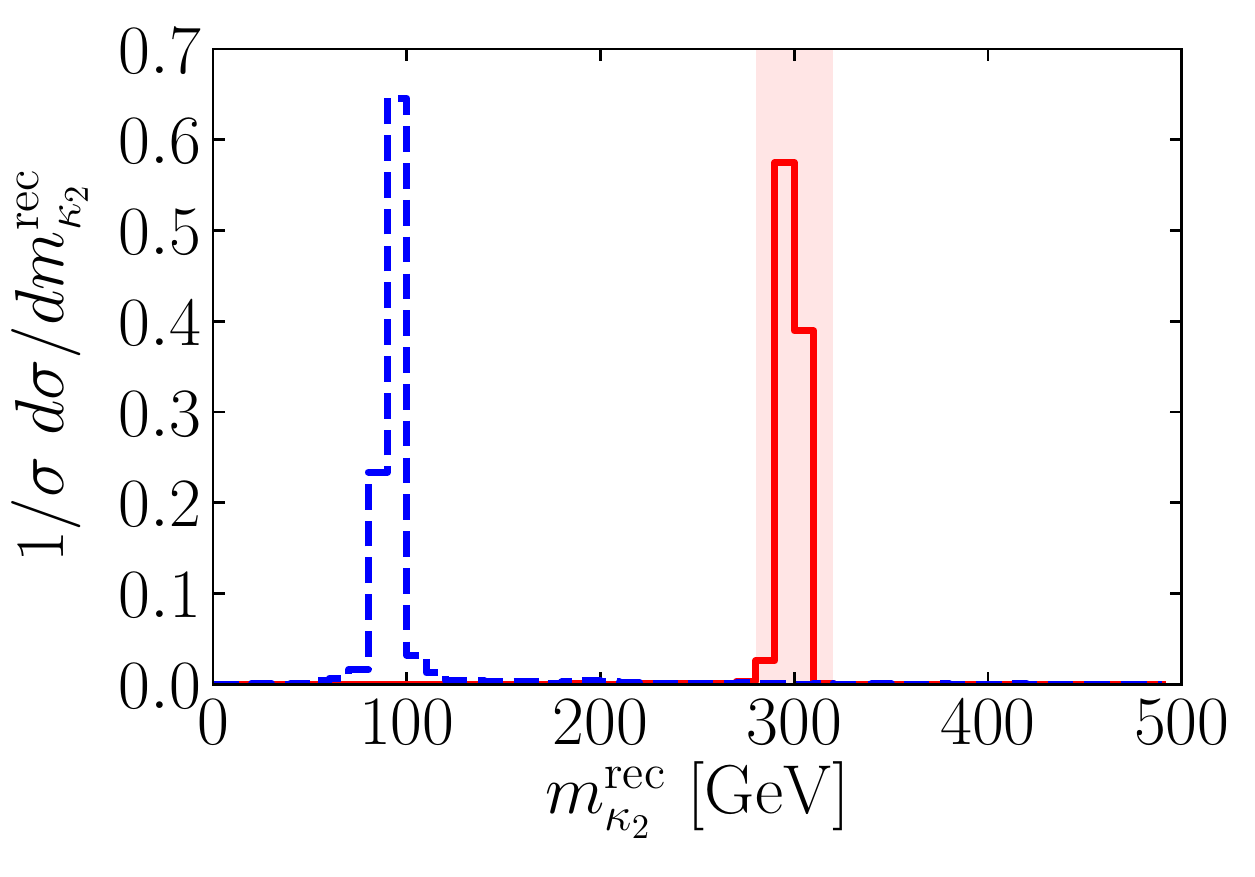}
  \includegraphics[width=0.49\columnwidth]{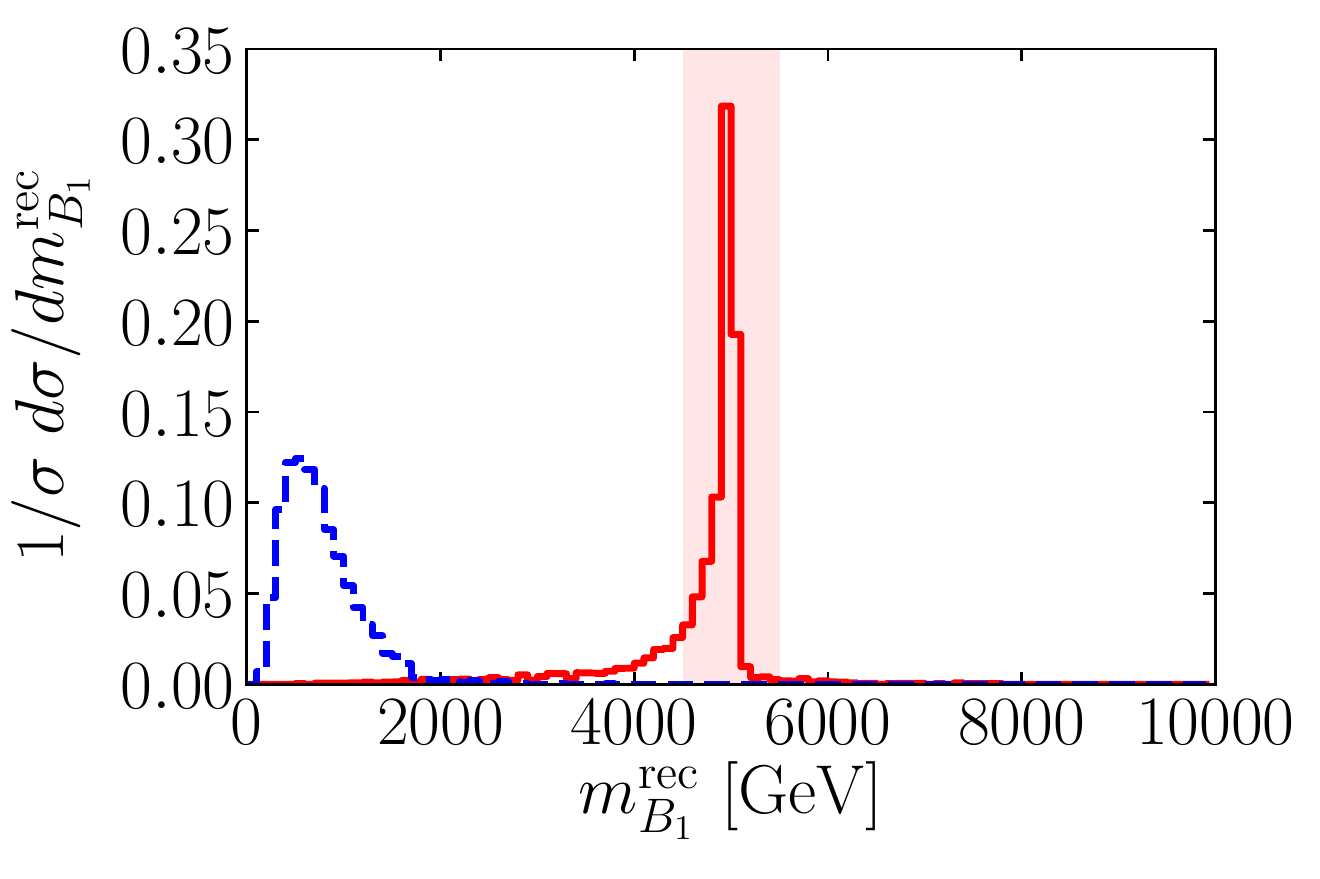}
  \includegraphics[width=0.49\columnwidth]{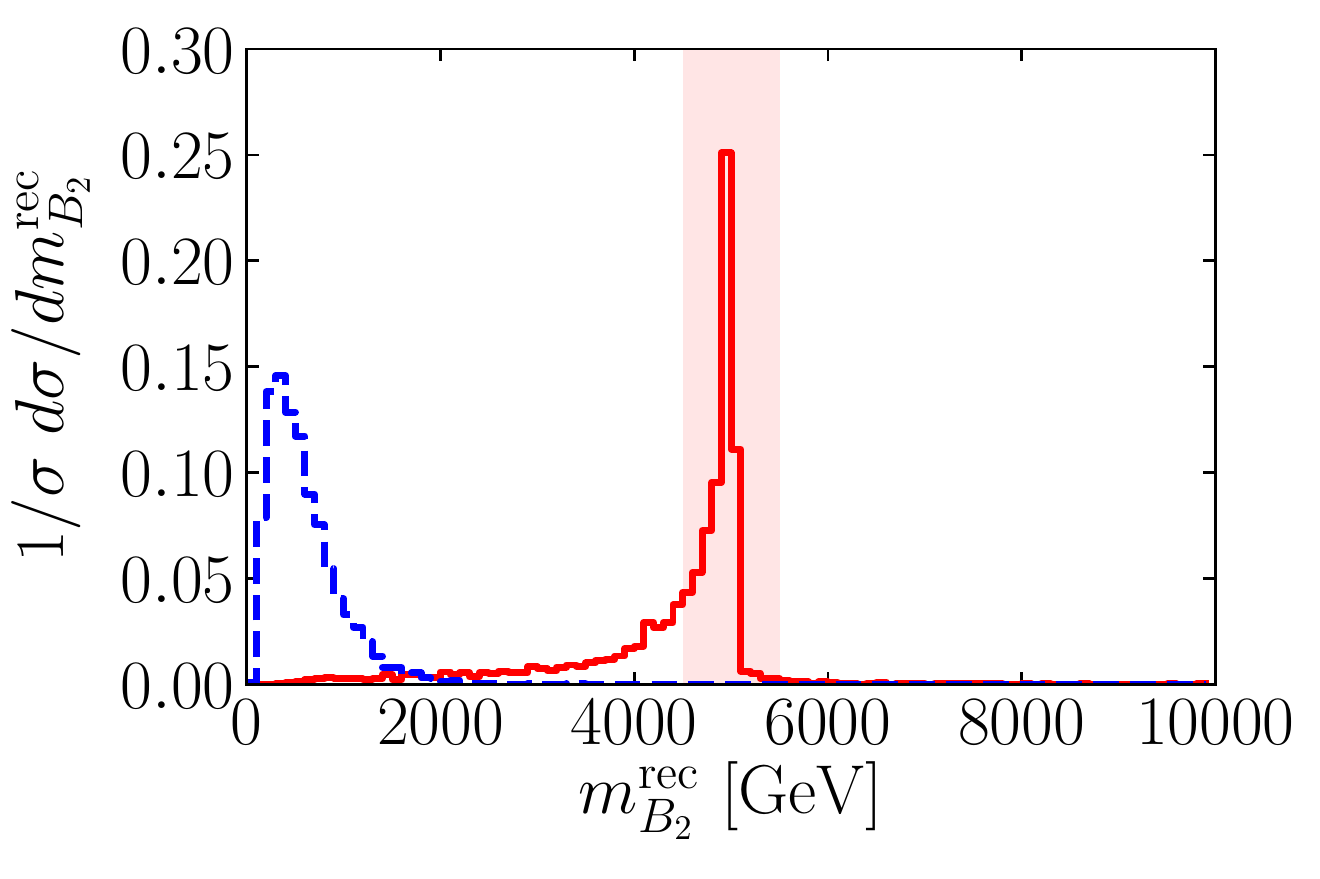}
 \end{center}
 \caption{\it Normalized distributions of $m_{\kappa_1}^\text{rec}$ (upper left), $m_{\kappa_2}^\text{rec}$ (upper right), $m_{B_1}^\text{rec}$ (bottom left) and $m_{B_2}^\text{rec}$ (bottom right) in the signal for $M = 5$ TeV and $m_\kappa = 300$ GeV (solid red) and in the SM background (dashed blue) in the analysis proposed for $pp\to B\overline{B}\to \kappa\kappa b\overline{b}$, $\kappa\kappa\to\mu^+ \mu^- \mu^+ \mu^-$.}
 \label{fig:distrVLQc}
\end{figure}

\begin{figure}[t]
 \includegraphics[width=0.49\columnwidth]{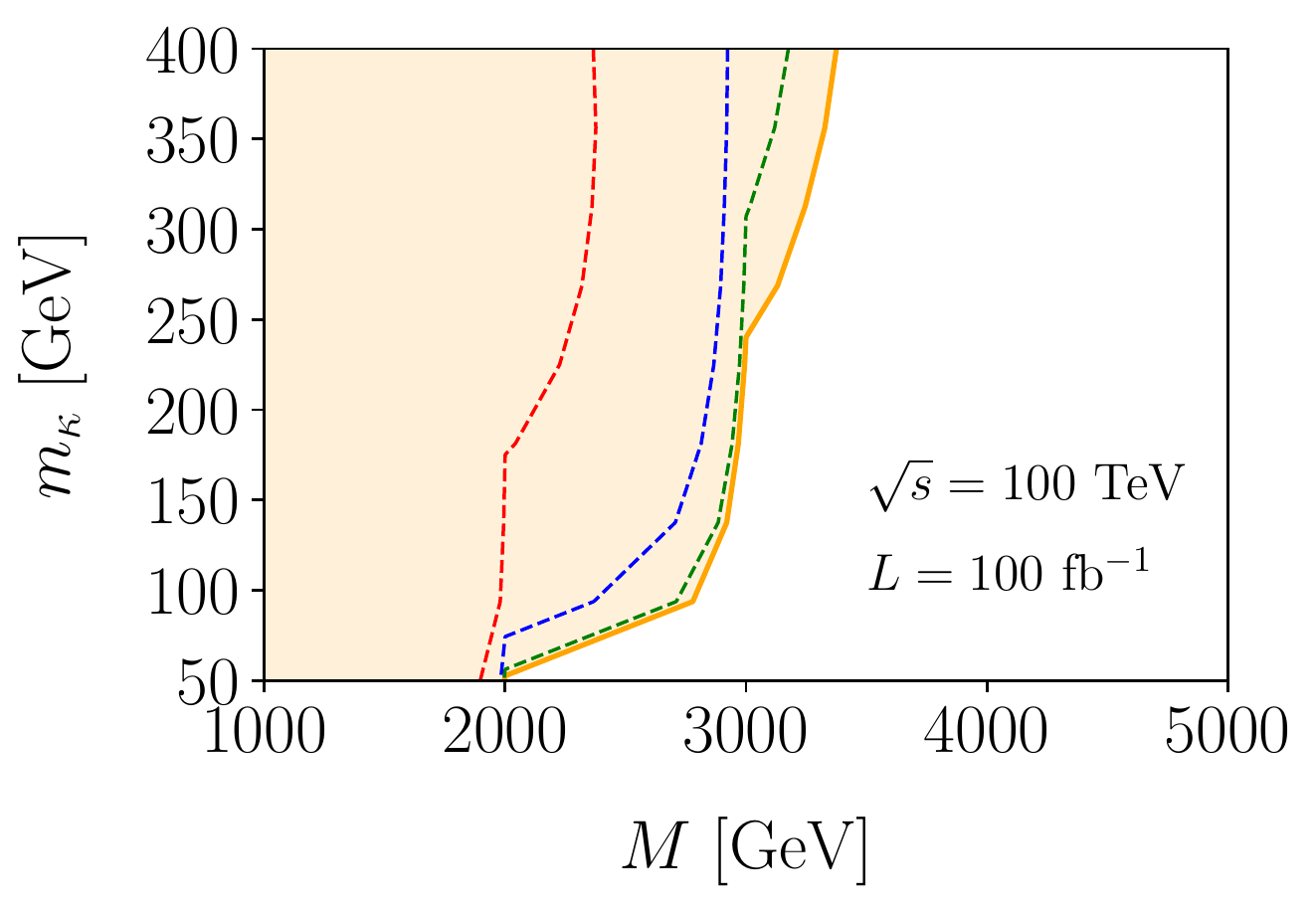}
 \includegraphics[width=0.49\columnwidth]{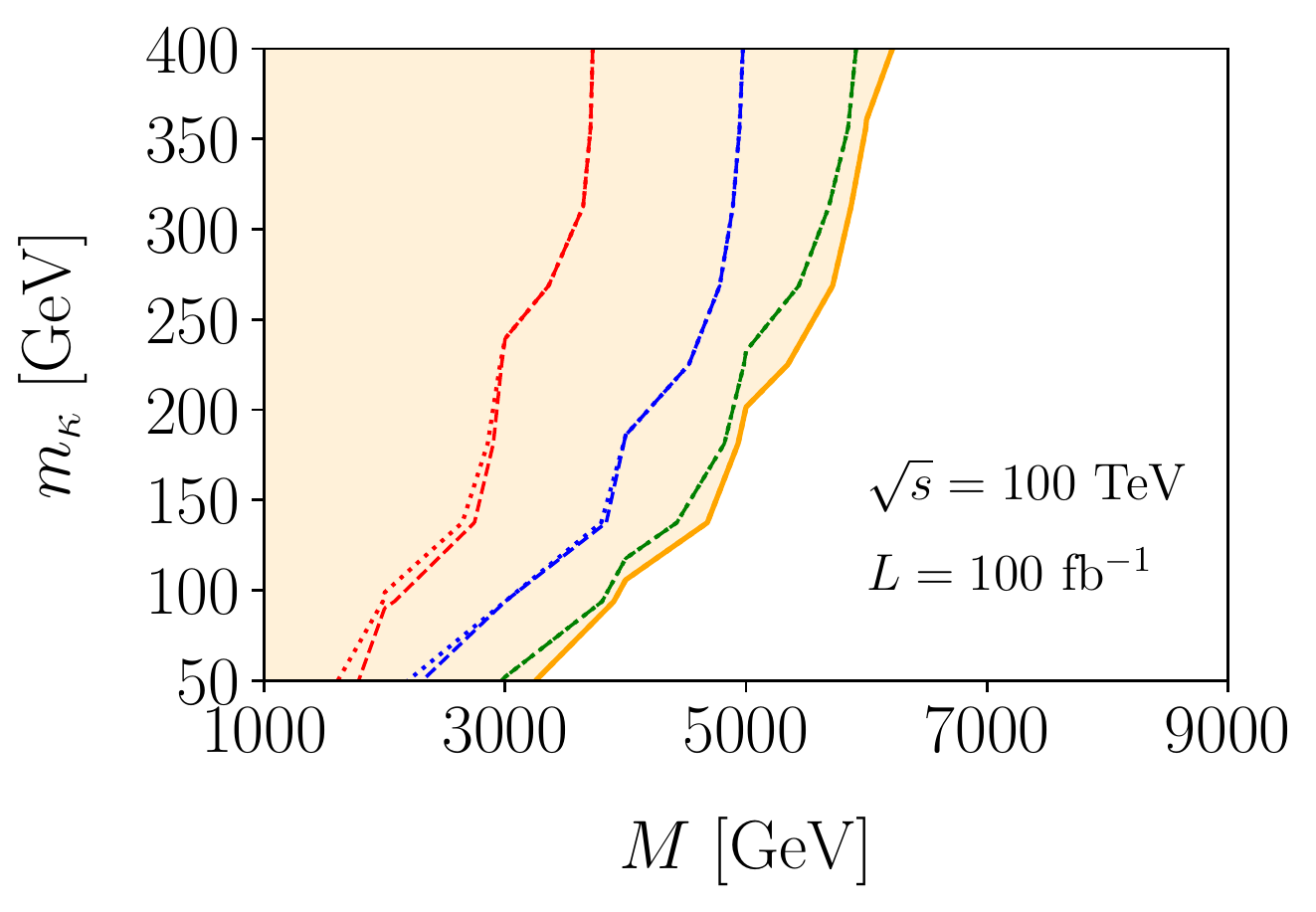}
 \caption{\it Exclusion lines at $95\%$ C.L. in the plane $(M, m_\kappa)$ for the first two analyses: $pp\to B\overline{B}\to \kappa\kappa b\overline{b}$, with $\kappa \kappa \rightarrow b \overline{b} \gamma \gamma$ (left) and $\kappa \kappa \rightarrow b \overline{b} b \overline{b}$ (right), assuming that the bottom partner decays $100\%$ into this final state. The results are presented for a future 100 TeV collider running with collected luminosity $L = 100~ {\rm fb}^{-1}$. The colored area enclosed by the orange solid, green dashed, blue dashed and red dashed lines can be excluded assuming ${\rm BR}(\kappa \rightarrow \overline{b} b) = 1,~0.8,~0.4$ and $0.1$, respectively. We have set ${\rm BR}(\kappa \rightarrow \gamma \gamma) = 10^{-3}$. In the right panel, the dashed (dotted) and solid lines assume a systematic uncertainty of 20\% (50\%) in the computation of the limits.
}\label{fig:VLQlimits}
\end{figure}

Nonetheless, the non-minimal structure of the model can be only established if, together with the process above,
decays of the VLQs into $\kappa$ are observed. No specific searches for this channel have been developed for 100 TeV colliders. Besides, only limits in a narrow range $100$ GeV $\lesssim m_\kappa\lesssim 400$ GeV have been obtained using LHC results for $\kappa\to b\overline{b}$~\cite{Chala:2017xgc}; see Refs. \cite{Anandakrishnan:2015yfa, Kraml:2016eti, Banerjee:2016wls, Aguilar-Saavedra:2017giu, Kim:2018mks, Alhazmi:2018whk, Kim:2019oyh, Dermisek:2019vkc, Benbrik:2019zdp, Cacciapaglia:2019zmj} for studies of other exotic decays of VLQs. Thus, we will provide dedicated searches for this case. We will focus on $pp\rightarrow B\overline{B}\to \kappa\kappa b\overline{b}$. 

There are a number of further reasons to concentrate
mostly on this channel. First, if the $SO(6)$ singlet is the lightest resonance, the DM channel is absent, while decays into $\kappa$ will be in general present, providing the only insight into the composite structure of the model. Second,
if the sextuplet is the lightest one, then $B''$ can decay mainly into $\kappa\, b$ 
\footnote{This conclusion differs from what was obtained in Ref. \cite{Serra:2015xfa}, under the assumption that the interactions of the vector-like resonance to the right-handed top quark, $c_R \overline{T_R}_i d_\mu^i \gamma^\mu t_R$ where $d_\mu^i \equiv -i {\rm Tr}[X^i U^\dagger D_\mu U]$, dominate over the interactions in equation \ref{eq:vlqdecay6}. On the contrary, our results are applicable in the scenario with $c_R \ll 1$.} as the coupling to the Higgs boson is suppressed by $\gamma$;
see equation~\ref{eq:vlqdecay6}. 
In our model, the requirement $\gamma \leq 1$ makes the branching ratio BR$(B\to h b)$ of the $SO(6)$ singlet significantly larger than BR$(B\to \kappa b)$. On the other hand, the exotic branching ratio is already $\gtrsim 50$\% for $\gamma \lesssim 0.4$ in the case of the sextuplet.
Third, after the decay of $\kappa$, this channel gives rise to a very clean signature, allowing to reconstruct the masses of $\kappa$ and $B$ relatively easily. Another benefit of this channel
is that it can be useful for CHMs not necessarily having DM, such as $SO(6)/SO(5)$. 
In addition, our analysis is also sensitive to pair-production of $T$, as demonstrated in following plots.

We will start by considering two decays of $\kappa\kappa$, namely $\gamma\gamma b\overline{b}$ and $b\overline{b} b\overline{b}$.
(These final states have
been proposed previously in studies of double Higgs production; see Refs.~\cite{Baur:2003gp, Grober:2010yv,  Baglio:2012np, Atre:2013ap, Barger:2013jfa, ATL-PHYS-PUB-2014-019, Azatov:2015oxa, ATL-PHYS-PUB-2017-001, Sirunyan:2017isc, Adhikary:2017jtu, Sirunyan:2018iwt}). In spite of the small branching ratio to two photons, ${\rm BR}\left(\kappa \rightarrow \gamma \gamma\right) \sim \alpha^2/(4\pi y_f)^2 m_\kappa^2/v^2$ with $y_f$ the Yukawa coupling of the fermion in the loop, 
the $\gamma\gamma b\overline{b}$ final state combines the large branching fraction of $k \rightarrow b \overline{b}$ and, at the same time, the low background and good mass resolution of the photon channel. Assuming a universal value for $c_\kappa$ in equation \ref{eq:Ly},
the final state with four bottom quarks has, in turn, the largest branching ratio for $m_\kappa \lesssim 2m_t$. Multi-jet searches in the context of non-minimal DM have been also studied in Ref. \cite{Dienes:2019krh}, aiming to probe multi-component dark sectors.

Finally, we will implement an analysis for $\kappa \kappa \rightarrow  \mu^+ \mu^- \mu^+ \mu^-$ which is intended to complement the very weak bounds from DM detection experiments that we have obtained for the leptophilic case.

\subsection{Search for $\kappa\kappa\to\gamma\gamma b\overline{b}$}
We generate signal and background events at $\sqrt{s} = 100$ TeV using \texttt{MadGraph v5}~\cite{Alwall:2014hca}. Signal
events are produced with no
parton level cuts. For the background, we consider the irreducible $pp\to b\overline{b} b\overline{b} \gamma\gamma$,
with $p_T^{b_1} > 500$ GeV and $p_T^\gamma > 20$ GeV, where $b_1$ stands for the leading $p_T$ $b$-quark. The cross section in this region of the parameter space is $\sim 0.006~{\rm pb}$. 
Signal and background events are subsequently showered using \texttt{Pythia v8}~\cite{Sjostrand:2014zea}. For the data
analysis, we use homemade routines based on \texttt{ROOT v6}~\cite{Brun:1997pa} and \texttt{Fasjet v3}~\cite{Cacciari:2011ma}. We disregard detector effects.

At the reconstruction level, a lepton is considered isolated if the hadronic activity within a cone of radius $\Delta R = 0.2$ is smaller than $10\%$ of its transverse momentum. Likewise for the photons.
We define jets using the anti-k$_t$ algorithm \cite{Cacciari:2008gp} with $R = 0.4$. Isolated leptons and photons with
$p_T < 20$ GeV and $|\eta| > 2.47$ (leptons) or $|\eta| > 2.5$ (photons)  are not included in the clustering process. We require exactly two isolated photons and at least four $b$-tagged
jets with $p_T > 20$ GeV. A jet is considered to be $b$-tagged if the angular separation between itself and a bottom meson is $\Delta R < 0.2$; the tagging efficiency being fixed to $0.7$. 
We also veto all events with a non-vanishing number of isolated leptons.

We name by $p_{\kappa_1}$ the four-momentum of the two photon system, its invariant mass being $m_{\kappa_1}^\text{rec}$. Likewise,
the momentum of the hadronically decaying scalar, $p_{\kappa_2}$, is considered to be the sum of that of the
two $b$-tagged jets with invariant mass closest to $m_{\kappa_1}^\text{rec}$. We will denote its invariant mass by
$m_{\kappa_2}^\text{rec}$.
There is a two-fold
ambiguity in assigning the remaining two hardest $b$-jets to either $\kappa_1$ or $\kappa_2$ to form the two vector-like $B$ quarks.
In order to solve it, we choose the combination
that minimizes $|m_{B_1}^\text{rec} - m_{B_2}^\text{rec}|$, with $m_{B_i}^\text{rec}$ the invariant
mass of the corresponding three-particle system.

In figure~\ref{fig:distrVLQa}, we show the normalized distributions of $m_{\kappa_1}^\text{rec}$, $m_{\kappa_2}^\text{rec}$, 
$m_{B_1}^\text{rec}$ and $m_{B_2}^\text{rec}$ in both the signal and the background, for a benchmark point defined by $M = 5$ TeV, $m_\kappa = 300$ GeV. The discriminating power of the observables is apparent. Thus, we select events fulfilling $|m_{\kappa_1}^\text{rec}-m_\kappa|<10$ GeV, $|m_{\kappa_2}^\text{rec}-m_\kappa|<100$ GeV, $|m_{B_1}^\text{rec}-m_B|<500$ GeV, $|m_{B_2}^\text{rec}-m_B|<500$ GeV. We compute the efficiency for selecting signal and background events for $M$ ($m_\kappa$) in the range $[1, 2,..., 9]$ TeV ($[50, 100,..., 400]$ GeV). Multiplying by the VLQ production cross section (ranging from $\sim 16$ to $8\times 10^{-5}$ pb), the scalar branching ratios and the luminosity, we get the final number of signal $s$ and background $b$ events after all cuts. For the selected mass windows, this analysis is basically background-free. 

We make use of the ${\rm CL}_s$ method \cite{Read:2002hq} for the calculation of exclusion limits. The maximum number of signal events is obtained by using the \texttt{TLimit} class of \texttt{ROOT}, considering a single bin analysis.  (We use a linear interpolation of the ${\rm CL}_s$ results to study the points that we did not scan.) The signal points which can be probed at the $95\%$ C.L. are shown in the left panel of figure~\ref{fig:VLQlimits} for an integrated luminosity of $L = 100~{\rm fb}^{-1}$. The regions enclosed by the solid orange line and the dashed green, blue and red lines can be probed at a future 100 TeV collider assuming ${\rm BR}(\kappa \rightarrow \overline{b} b ) = 1$ and $0.8,~0.4$ and $0.1$, respectively. We have fixed ${\rm BR}(\kappa \rightarrow \gamma \gamma) = 10^{-3}$.

\subsection{Search for $\kappa\kappa\to b \overline{b} b\overline{b}$}
Signal events are again produced with no parton level cuts. For the background, we consider $pp\rightarrow b\overline{b} b\overline{b}$, with $p_T^{b_1} > 500$ GeV. 
The extra two $b$-jets are found among the products of the Pythia shower.
Both $pp\rightarrow b\overline{b} b \overline{b}$ and $pp\rightarrow b\overline{b} b\overline{b} b\overline{b}$ have very similar distributions, validating the use of the former as the background sample\footnote{We will further assume systematic uncertainties in the background, for the computation of the limits, to infer the impact of this assumption.}. This sample is normalized to the cross section of the actual background as computed with \texttt{MadGraph}, roughly $\sim 13$ pb in this region of the parameter space. 

We use the same lepton and jet definitions and isolation criteria as before. We require no isolated leptons and at least six $b$-tagged jets.
The $b$-tagged jets reconstructing the two $\kappa$'s are those minimizing $|m_{\kappa_1}^\text{rec}-m_{\kappa_2}^\text{rec}|$ among the four with smallest $p_T$, where $m_{\kappa}^\text{rec}$ stands now for the invariant mass of two $b$-jets. Again, there is a two-fold ambiguity in assigning the two hardest jets to any of the $\kappa$'s to form $B_1$ and $B_2$; it is solved in the same way as before. 

The normalized distributions of $m_{\kappa_1}^\text{rec}$, $m_{\kappa_2}^\text{rec}$, 
$m_{B_1}^\text{rec}$ and $m_{B_2}^\text{rec}$ in both the signal and the background for the same benchmark point are shown in figure~\ref{fig:distrVLQb}. In order to show that the channel $pp\to T\overline{T} \to \kappa \kappa t \overline{t}$ is also captured by this analysis, we plot the corresponding distributions in figure~\ref{fig:distrVLQTPb}.

We require all events to have $|m_{\kappa_1}^\text{rec}-m_\kappa|<100$ GeV, $|m_{\kappa_2}^\text{rec}-m_\kappa|<100$ GeV, $|m_{B_1}^\text{rec}-m_B|<500$ GeV, $|m_{B_2}^\text{rec}-m_B|<500$ GeV. (For the benchmark point, the top partner signal efficiency is about $1/3$ of the bottom partner signal efficiency.) Performing a statistical analysis equivalent to the one used in the last section, we obtain the plot in the right panel of figure~\ref{fig:VLQlimits}. The dashed (dotted) lines are obtained for a systematic uncertainty of 20\% (50\%) in the number of background events after all cuts. Given that the analysis is nearly background-free for $M \gtrsim 3$ TeV, the latter has a minor impact in the results. 

\subsection{Search for $\kappa\kappa\to \mu^+ \mu^- \mu^+ \mu^-$}
Parton level cuts are only applied for the irreducible background, $p p \rightarrow b \overline{b} \mu^+ \mu^- \mu^+ \mu^-$, with $p_T^{b_1} > 500$ GeV. The corresponding cross section is $\sim 10 ^{-5}~{\rm pb}$.
We use the same lepton and jet definitions and isolation criteria as before. We require no isolated electrons, two pairs of muons with opposite charge and exactly two $b$-tagged jets. 

The muon pairs reconstructing the two $\kappa$'s are those minimizing $|m_{\kappa_1}^{\rm{rec}} - m_{\kappa_2}^{\rm{rec}}|$, where $m_{\kappa}^\text{rec}$ stands for the invariant mass of two oppositely charged muons.
To decide which $b$-jet is assigned to each $\kappa$, we choose again the combination that gives the minimum difference between the reconstructed heavy quark masses.

The normalized mass distributions in both the signal and the background are plotted in figure \ref{fig:distrVLQc} for the benchmark point considered in the previous sections.
We require all events to have $|m_{\kappa_1}^\text{rec}-m_\kappa|<20$ GeV, $|m_{\kappa_2}^\text{rec}-m_\kappa|<20$ GeV, $|m_{B_1}^\text{rec}-m_B|<500$ GeV, $|m_{B_2}^\text{rec}-m_B|<500$ GeV. 
After this last cut, the analysis becomes essentially background-free. 
For different VLQ masses, the signal efficiency is nearly constant ($\epsilon_s \approx 0.1-0.2$); therefore, the exclusion lines in the $(M, m_\kappa)$ plane are close to vertical. All masses up to $M \sim 5~(8)$ TeV are excluded at $95\%$ C.L. for ${\rm BR}(\kappa \rightarrow \mu^+ \mu^-) = 0.1~(0.4)$, at a future 100 TeV collider with integrated luminosity $L = 1~{\rm ab}^{-1}$.

\section{Conclusions}\label{sec:conclusions}

DM from CHMs has become an extensive topic of research in recent years. Besides the minimal model, delivering only one stable scalar besides the Higgs boson, different non-minimal constructions have been discussed to which DM constraints apply differently. Previous works have assumed that the \textit{DM is the lightest} particle in the spectrum, so that the extra degrees of freedom are decoupled from its phenomenology. This assumption does not always hold, and strong constraints from DM experiments can actually be evaded efficiently in such case.

We have studied this setup in the next-to-minimal anomalous-free $SO(7)/SO(6)$ CHM, where an extra singlet $\left(\kappa\right)$ besides DM $\left(\eta\right)$ arises in the pNGB spectrum, opening a new annihilation channel for DM. We have confronted two cases, justified by the different embeddings of the SM quarks: one in which the DM coupling to the Higgs boson at zero momentum is sizable while the zero-momentum coupling to the extra scalar is absent (RegI); and a second one where this hierarchy is inverted (RegII).  

We have studied the phenomenology of both regimes,  
for $100 \leq m_\eta < 1500$ GeV and $m_\kappa \leq m_\eta$. In RegI, the annihilation fraction to the new scalar is sizable ($\gtrsim 30 \%$) in a wide range of the parameter space. This is due to the partial cancellation in the SM annihilation channel between the scalar portal coupling and new derivative interactions originating from the new physics sector. In this regime, we also find that the relic density constraint excludes confining scales ${f \gtrsim 3.3~\rm{TeV}}$, keeping the model \textit{natural}. While a wide region of the parameter space is unconstrained by XENON1T, this regime can be entirely probed by the future LZ experiment.
On the other hand, in RegII, the scalar contributions for the nucleon--DM cross section are suppressed by loop factors. Therefore, this regime evades all current and future direct detection constraints.

To further test the model, we have studied the constraints from indirect detection searches. We have derived new gamma-ray bounds resulting from dedicated simulations of the exotic DM annihilation into $\kappa\kappa$, relying on the Fermi-LAT likelihoods released by the collaboration.
In particular, the decay channels into light leptons are very weakly constrained. 
Even if future experiments strengthen these bounds by an order of magnitude, a thermal relic with mass $m_\eta \approx 200$ GeV that decays mainly into a muonphilic $\kappa$ will still be unconstrained while accounting for all the observed DM (in RegII).

Finally, we have developed dedicated collider analyses to complement the DM probes. We presented novel strategies to search for pair produced third generation vector-like quarks that decay to their SM partner and the extra singlet. This channel allows us to test both the compositeness and non-minimality of the model and it is generally present, whereas the DM channel is not. We focused on three decays of the $\kappa$-particle, leading to $b \overline{b} \gamma \gamma$, $b \overline{b} b\overline{b}$ and $2\mu^+ 2\mu^-$ final states. From the first two analyses, we conclude that VLQ masses as large as $M \sim 3$ and 6 TeV, respectively, can be tested at a future $100$ TeV collider with collected luminosity $L = 100~{\rm fb}^{-1}$.
Performing the third analysis, we can probe masses up to $m_\kappa \sim 400$ GeV, $M \sim 9$ TeV with $L = 1 ~\rm{ab}^{-1}$, provided that $\kappa$ decays mainly to muons.
These searches can be our only insight into the model in regimes that escape all constraints from the DM experiments, even at future facilities.

Altogether, notwithstanding the growing efforts in building alternative models of DM, our study indicates that thermal WIMPs can be viable candidates in non-minimal --- but still \textit{natural} --- composite Higgs scenarios.

\section*{ Acknowledgments}
\noindent
I am particularly grateful to Mikael Chala for fruitful discussions that motivated this work, valuable feedback on the results and useful comments on the manuscript. It is a pleasure to thank the MITP for the hospitality during the completion of this work. I would like to thank Nuno Castro, Alexander Belyaev, Jan Heisig and Yann Gouttenoire for enlightening discussions. I would also like to thank Guilherme Guedes and Tiago Vale for reviewing parts of the manuscript.
This work is supported by Funda\c{c}\~ao para a Ci\^encia e Tecnologia (FCT) under the grant PD/BD/142773/2018 and LIP (FCT, COMPETE2020-Portugal2020, FEDER, POCI-01-0145-FEDER-007334).

\noindent

\clearpage
\bibliographystyle{style}
\bibliography{notes}{}

\end{document}